\documentclass[review,onefignum,onetabnum]{siamonline250211}

\usepackage{subfigure}
\usepackage{amssymb}

\usepackage{pdfpages}

\usepackage{lipsum}
\usepackage{amsfonts}
\usepackage{graphicx}
\usepackage{epstopdf}
\usepackage{algorithmic}
\ifpdf
  \DeclareGraphicsExtensions{.eps,.pdf,.png,.jpg}
\else
  \DeclareGraphicsExtensions{.eps}
\fi

\usepackage{enumitem}
\setlist[enumerate]{leftmargin=.5in}
\setlist[itemize]{leftmargin=.5in}

\newsiamremark{remark}{Remark}
\newsiamremark{hypothesis}{Hypothesis}
\crefname{hypothesis}{Hypothesis}{Hypotheses}
\newsiamthm{claim}{Claim}
\newsiamremark{fact}{Fact}
\crefname{fact}{Fact}{Facts}

\headers{Influence of topological and geometric disorder on the spectrum of the differential Laplacian on networks}{C. E. Maher, J. L. Marzuola, and K. A. Newhall}

\title{Probing the influence of topological and geometric disorder on the spectrum of the differential Laplacian operator on networks\thanks{Submitted to the editors \today.
\funding{CEM is supported by the collaborative NSF DMREF Grant No. CMMI-2323342. KAN is supported by the collaborative NSF DMREF Grant No. CMMI-2323342 and NSF Grant No. DMS-2307297.  JLM is supported by NSF Applied Math Grant DMS-2307384.}}}

\author{Charles Emmett Maher\thanks{Department of Mathematics, University of North Carolina at Chapel Hill, Chapel Hill, NC, USA 
  (\email{cemaher@email.unc.edu},\email{marzuola@email.unc.edu},\email{knewhall@email.unc.edu}).}
\and Jeremy L. Marzuola\footnotemark[2]
\and Katherine A. Newhall\footnotemark[2]}

\usepackage{amsopn}

\ifpdf
\hypersetup{
  pdftitle={Probing the influence of topological and geometric disorder on the spectrum of the differential Laplacian operator on networks},
  pdfauthor={C. E. Maher, J. L. Marzuola, and K. A. Newhall}
}
\fi

\begin{document}

\maketitle

\begin{abstract}
Metric networks are network-shaped, one-dimensional structures on which one can solve differential equations to simulate a wide range of physical systems including conjugated molecules, photonic crystals, quantum mechanics in waveguide networks, and acoustic metamaterials. 
More concretely, a metric network is a network whose edges are each assigned a notion of length and a coordinate describing position.
One can then define function spaces and differential operators on these objects to model the aforementioned systems. 
Recent software advancements have made it feasible to analyze partial differential equations on large, compact metric networks with a vast array of structures. 
Here, we generate compact metric network structures using the spatial tessellations of two-dimensional hyperuniform point patterns, which have suppressed large-scale density fluctuations relative to typical disordered point patterns. 
This choice of structure is inspired by the exotic physical properties of network materials with these structures in other contexts. 
Then, we characterize the eigenvalue spectrum structure of the differential Laplace operator on these networks. 
In particular, we find that gaps can form in the eigenvalue spectra of these networks whose widths increase when the distribution of edge lengths is narrow and as the number of triangular faces increases.
Importantly, many of the structures we consider are realizable in Euclidean space, meaning they are well-suited for practical applications in, e.g., metamaterial design. This work can thus be used to inform the design of metric network-based systems with spectral gaps with tunable widths and locations.

\end{abstract}



\section{Introduction}
Network structures are of fundamental importance in a plethora of contexts including biological systems \cite{pujol_impact_2012, wang_mechanosensitive_2025}, industrial applications \cite{fugenschuh_chapter_2015, paul_analysis_2007}, and pure mathematics \cite{barthelemy_spatial_2022}.
Of present interest is the application of networks to materials science \cite{mcinerney_topological_2025, mendels_systematic_2023, obrero_electrical_2025, picu_network_2022, areyes-martinez_tuning_2022, siedentop_stealthy_2024}.
Disordered network metamaterials are a material class of particular interest due to their wide variety of exotic physical properties including negative Poisson's ratio \cite{rreid_ideal_2019,reid_auxetic_2018, shen_autonomous_2024}, exotic mechanical responses \cite{Surjadi2025,Zhang2024}, and the ability to mimic the mechanical properties of biological tissues \cite{mohammed_design_2018, ramirez-torres_three_2018, sniechowski_heterogeneous_2015,  wit_simulation_2019}.
Recent advancements in additive manufacturing methods have made realizing computationally-generated network structures feasible, allowing for the direct experimental assessment of the theoretically-predicted properties of such structures \cite{moody_methodology_2025}.
With these experimental developments, it is increasingly important to understand the relationship between the structure of a network and its properties so we can manufacture materials with tailored physical properties.
Such materials have applications in diverse fields including aerospace \cite{alderson_auxetic_2007} and biomedical devices \cite{carneiro_auxetic_2013}.
Importantly, in each of the areas noted so far, it is important to consider the entire spatial extent of the network structure.

Metric networks\footnote{We note that these objects are also often referred to as metric \textit{graphs}, but we opt to use the term metric network to be consistent with other mentions of network objects in this paper.}, which can be thought of as network-shaped spaces, are a natural way to describe these spatially extended network systems where the dynamics of interest take place on the edges \cite{berkolaiko_introduction_2013}.
More specifically, a metric network consists of a set of vertices and edges where each edge is associated with a positive length and a coordinate describing a position on the edge.
One can then define function spaces and differential operators on the metric network (a combination sometimes called a quantum graph) and compute the solutions to differential equations on these networks. 

A wide range of physical systems can be modeled with this framework, including molecules \cite{atilgan_anisotropy_2001, bistritzer_moire_2011, doruker_dynamics_2000, kuchment_graph_2002}, photonic crystals \cite{axmann_asymptotic_1999, kuchment_differential_2002, kuchment_spectral_1999}, quantum mechanics in waveguide networks \cite{dietz_closed_2024}, and acoustic metamaterials \cite{lawrie_application_2024}.
By contrast, the common notion of combinatorial networks, which consist of a set of vertices that are connected by a set of edges that describe the relationships between discrete entities in the network \cite{newman_networks_2018}, are used to model, e.g., ecological systems \cite{emmert-streib_gene_2014}, metabolic \cite{jeong_large-scale_2000} and gene regulatory networks \cite{brose_embedding_2025}, and opinion dynamics \cite{starnini_opinion_2025}.
Because there is no notion of space on combinatorial networks, one can only model dynamics on them using ordinary differential equations.
The dynamics on combinatorial networks are often described using the graph Laplacian matrix \cite{mcgraw_laplacian_2008}, which is a discrete analog of the differential Laplacian that can be used to model wave and heat transport on metric networks.

A problem of interest for metric networks is the characterization of the eigenvalue spectrum of the differential (i.e., not graph) Laplacian operator (see, e.g., Chapter 3 of Ref. \cite{berkolaiko_introduction_2013} and references therein for an overview).
Of particular interest are gaps in this spectrum, which are important to the study of solid-state physics \cite{kuchment_spectra_2007}, photonics \cite{kuchment_spectral_1999}, and waveguide systems \cite{lawrie_quantum_2022, yoshitomi_band_1998}.
A large amount of the work on gaps in the spectra of metric networks focuses on the spectral gap, which is the gap between the 0 eigenvalue and the next largest one \cite{band_quantum_2017, berkolaiko_impediments_2023, borthwick_sharp_2021, kennedy_spectral_2016, khrabustovskyi_periodic_2020, kurasov_spectral_2013}.
In periodic metric networks, one can create gaps elsewhere in the eigenvalue spectrum by applying ``decorations'' to the vertices, following the method from Ref. \cite{berkolaiko_introduction_2013} which adapts a method used for combinatorial networks \cite{schenker_creation_2000}.
The study of differential equations on metric networks has focused primarily on small periodic and \textit{compact} networks \cite{porter_nonlinearity_2019}, which have a finite number of edges that have finite lengths.
Relatively little work has been done on large compact metric networks due to the inaccessibility of methods to accurately compute solutions to differential equations on such structures.

Recent advances in computing have made it feasible to study the spectra of and simulate dynamics on large, finite metric network structures.
For example, Holden and Vasil developed code to compute the spectrum of the Laplacian on large finite and periodic metric networks and showed that the solutions of partial differential equations (PDEs) on metric networks embedded in a manifold converge to the solution of its corresponding PDE posed on the manifold itself as the density of vertices increases \cite{holden_continuum_2025}.
B\"ottcher and Porter have developed a Python package that treats a large class of PDEs on metric networks with up to $\sim\!10^4$ edges and $\sim\!10^4$ vertices to probe the relationship between structure and dynamics on metric networks \cite{bottcher_dynamical_2025}.
Weller has developed a similar Julia package for elliptic and parabolic PDEs on metric networks \cite{weller_numerical_2024}.
Here, we use QGLAB, a MATLAB package developed by Goodman, Conte, and Marzuola that is used for computing the spectra of elliptic operators and simulating the time evolution of PDEs on metric networks \cite{goodman_qglab_2025}.
We choose to use QGLAB due to its adaptability in constructing large random metric networks and building the corresponding Laplacian operator \cite{goodman_qglab_2025}.

To narrow down the space of possible network structures to examine, we consider networks generated from the spatial tessellations of hyperuniform point patterns, which are of interest due to their broad range of desirable physical properties \cite{chen_topological_2021, chen_stonewales_2021, florescu_designer_2009, jiao_exotic_2025, man_isotropic_2013, siedentop_stealthy_2024}.
Hyperuniform point patterns have anomalously suppressed density fluctuations compared to those in typical disordered systems \cite{torquato_local_2003}, resulting in a structure factor $S(\mathbf{k})$ that tends to 0 as the wavenumber $k\equiv|\mathbf{k}|$ tends to 0.
Of particular interest are disordered stealthy hyperuniform systems, whose structure factors are identically zero for some set of wavenumbers around the origin \cite{torquato_ensemble_2015}.
Network materials generated from the tessellations of these point patterns have desirable physical properties, for example, nearly optimal elastic moduli and thermal and electrical conductivity \cite{skolnick_effective_2025, torquato_multifunctional_2018}.
In addition, these network materials can possess complete, isotropic photonic bandgaps \cite{florescu_designer_2009, man_isotropic_2013, siedentop_stealthy_2024}.
Inspired by these exotic properties, here we generate metric networks from the tessellations of hyperuniform point patterns and characterize their eigenvalue spectra.

In this work, we generate compact metric network structures from the tessellations of disordered two-dimensional (2D) hyperuniform, disordered stealthy hyperuniform, and nonhyperuniform point patterns with different degrees of short-scale translational disorder.
In particular, we apply the Delaunay tessellation \cite{barthelemy_spatial_2022}, Gabriel tessellation \cite{gabriel_new_1969}, and Delaunay-centroidal tessellation \cite{florescu_designer_2009} schemes to these point patterns respecting periodic boundary conditions, then break the periodicity in different ways, which are discussed in detail in Sec. \ref{Sec:Methods}.
We focus primarily on compact metric network systems due to the existing capabilities to manufacture such structures in a laboratory setting \cite{moody_methodology_2025}. 
Then, using QGLAB, we compute the eigenvalues and eigenmodes of the combinations of the point patterns, tessellation schemes, and boundary types above and examine how they affect the eigensystem.
We find that 2D metric networks with a broader distribution of edge lengths and fewer triangular cells tend to have flatter eigenvalue distributions.
By contrast, metric networks with a narrow distribution of edge lengths and a large fraction of triangular cells exhibit deep troughs in their eigenvalue distributions.
These results offer a method to tune the position and size of gaps in the eigenvalue spectrum of disordered compact metric networks.
Moreover, because eigenmodes can be used to construct solutions of linear PDEs, this study of the eigensystems of metric networks can be used to inform the design of disordered network metamaterials with, e.g., targeted heat and wave transport characteristics.

Spectral properties of disordered metric networks have been studied in a variety of ways previously, but not regularly at the scale of the models generated here and with comparable disorder.  
For instance, in Ref. \cite{gnutzmann_quantum_2006}, the authors consider the large-network limit of spectral average functions and spectral two-point correlation functions.  They consider both a periodic-orbit approach and a supersymmetry approach, with the disorder generated by a fixed choice of rationally independent bond lengths.  
The authors give many details on star networks, but discuss how to apply spectral statistics more broadly in their setting.  
Similarly, in Ref. \cite{gnutzmann2010eigenfunction} they consider the autocorrelation statistics of high-energy eigenfunctions in large metric networks. 
In Refs. \cite{alon2018nodal} and \cite{alon2024universality}, the authors consider a family of disordered networks and study the statistics of the distribution of the oscillation of eigenvectors relative to the maximal amount of oscillation possible (as related to the energy of the eigenfunction and the number of cycles in the network).  
On trees for instance, as the eigenvalue increases in energy, the eigenfunction increases in oscillation.  
However, in more topologically interesting examples, this is a much more challenging question, and the authors prove that the oscillations distribute in a Gaussian fashion along the allowed values for a variety of networks in a large network limit.  
In Ref. \cite{silva_quantum_2024}, the authors consider the related problem of randomized transmission/reflection coefficients for disordered networks attached to incoming about outgoing leads.  
The total number of vertices in the networks studied however were relatively small. 
The authors in Ref. \cite{kottos2001quantum} consider the scattering matrix in metric networks with both randomly chosen edges and randomly determined vertex conditions.

The rest of this paper is organized as follows.
In Sec. \ref{Sec:Background}, we present the pertinent mathematical background and definitions.
In Sec. \ref{Sec:Methods}, we describe how we generate the metric networks examined in this work and how we characterize their eigensystems.
In Sec. \ref{Sec:Results}, we characterize the structure and eigensystems of metric networks and examine the effects of disorder on their eigenvalue spectra.
Finally, in Sec. \ref{Sec:Conclusions} we offer concluding remarks and outlook for future studies. 

\section{Mathematical Background} \label{Sec:Background}
In this section, we first review the construction of metric networks and how one can define differential operators on them, following closely the setup in Ref. \cite{goodman_qglab_2025}.
Then, we present the basic definitions associated with hyperuniform point patterns and networks.

\subsection{Metric networks}
A metric network can be considered as a complex of edges on which one can define function spaces and differential operators.
First, consider a directed combinatorial network $\Gamma=(\mathcal{V},\mathcal{E})$ with a set of vertices $\mathcal{V}=\{\mathtt{v}_n,n=1,\dots,|\mathcal{V}|\}$ and set of directed edges $\mathcal{E}=\{\mathtt{e}_m=(\mathtt{v}_i\rightarrow\mathtt{v}_j),m=1,\dots,|\mathcal{E}|\}$.
The degree of each vertex $d_i$ is the number of edges that either originate or terminate at vertex $\mathtt{v}_i$.
To obtain a metric network, we associate a real interval $[0,\ell_{m}]$ of length $\ell_m$ to each edge and a coordinate $x$ that increases from 0 to $\ell_m$ as the edge is traversed from $\mathtt{v}_i$ to $\mathtt{v}_j$.
Here, we only consider compact metric networks, so $|\mathcal{V}|$, $|\mathcal{E}|$, and all $\ell_m$ are positive and finite.

A function $\mathbf{\Psi}(x)$ defined on $\Gamma$ can be interpreted as a collection of functions defined on each of the edges of $\Gamma$, $\mathbf{\Psi}|_{\mathtt{e}_m}=\Psi_m(x)$.
A Laplace operator on $\Gamma$ is defined by
\begin{equation}
    \bigtriangleup|_{\mathtt{e}_m}=\frac{\textrm{d}^2}{\textrm{d}x^2},\quad\textrm{for}\quad0<x<\ell_m,
\end{equation}
subject to the boundary conditions imposed at the vertices.
Here, we employ Neumann--Kirchhoff boundary conditions at each vertex and Dirichlet boundary conditions on the metric network itself.
This set of boundary conditions results in a self-adjoint Laplacian, though there are other possible boundary conditions that would yield a self-adjoint Laplacian (see, e.g, Ref. \cite{berkolaiko_introduction_2013}).
The first $d_n-1$ of the $d_n$ equalities needed to define the vertex condition at $\mathtt{v}_n$ impose continuity at the vertex. 
Let $\mathcal{V}_n$ be the set of edges adjacent to $\mathtt{v}_n$.
These $d_n-1$ equalities allow us to define $\mathbf{\Psi}(\mathtt{v}_n)$ by

\begin{equation}
    \label{eq:VertCont}\mathbf{\Psi}(\mathtt{v}_n)\equiv\psi_i(\mathtt{v}_n)=\psi_j(\mathtt{v}_n),\forall\mathtt{e}_i,\mathtt{e}_j\in\mathcal{V}_n.
\end{equation}
The $d_n$th equality at $\mathtt{v}_n$, the Neumann--Kirchhoff flux condition, is then
\begin{equation}
\label{eq:NK}\sum_{\mathtt{e}_m\in\mathcal{V}_n}\psi_m'(\mathtt{v}_n)=0,
\end{equation}
where the derivative is always in the direction pointing away from the vertex.
We note that this choice of boundary conditions also has a physical interpretation, namely these are the natural boundary conditions to describe the behavior of waves in a connected network of cylindrical waveguides in the limit that the waveguide radius vanishes.

Now, with these vertex and boundary conditions, we can define a suitable norms and functions spaces for the Laplacian.
In particular, $L^p(\Gamma)$:
\begin{equation}
||\mathbf{\Psi}||^p_{L^p(\Gamma)}=\sum^{|\mathcal{E}|}_{m=1}||\psi_m||^p_{L^p},
\end{equation}
and the $L^2$ inner product.
\begin{equation}
\label{eq:L2IP}\langle\mathbf{\Psi},\mathbf{\Phi}\rangle=\sum^{|\mathcal{E}|}_{m=1}\int_0^{\ell_m}\psi^*_m(x)\phi_m(x)\textrm{d}x.
\end{equation}
In addition, we define $H^1(\Gamma)$ as the space of square-integrable functions with square-integrable first derivatives and $H^2(\Gamma)$ that additionally has square-integrable second derivatives. 
Specifically, we define $L^2(\Gamma)$, $H^1(\Gamma)$, and $H^2(\Gamma)$ as the space of functions that satisfy the conditions above edge-by-edge.
In addition, we define $L^2_\Gamma$ as the space $L^2(\Gamma)$ equipped with the inner product in Eq. (\ref{eq:L2IP}), $H^1_\Gamma$ as the space of functions in $H_1(\Gamma)$ that also satisfy the continuity equations in Eq. (\ref{eq:VertCont}) and $H_\Gamma^2$ as the set of functions in $H^2(\Gamma)$ that satisfy Eqs. (\ref{eq:VertCont}) and (\ref{eq:NK}).

\begin{figure}
    \centering
    \includegraphics[width=0.5\linewidth]{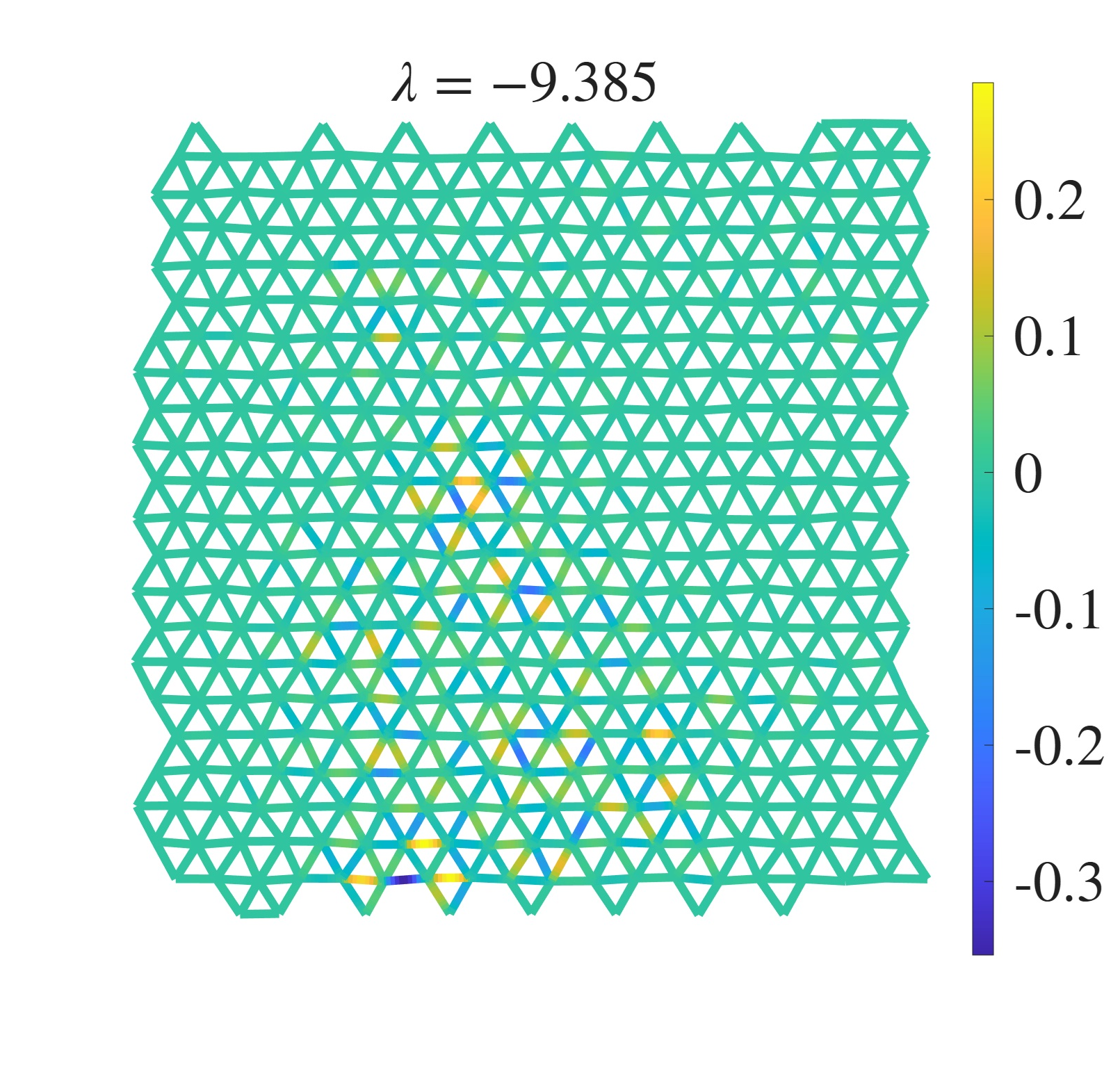}
    \caption{An example eigenmode on one of the disordered metric network structures considered herein. This network is generated from the Delaunay tessellation of a hyperuniform $A_2$ URL point pattern with $N = 418$ particles, $a = 0.1$, and prune boundary conditions (all of which are defined in Sec. \ref{Sec:Methods}). This eigenmode with eigenvalue $\lambda=-9.385$ corresponds to one in the ``central island'' described in Sec. \ref{Sec:Results}.}
    \label{fig:configex}
\end{figure}

With this Laplacian operator and suitable function spaces, we then consider the spectrum and corresponding eigenmodes of the Laplacian defined on a given $\Gamma$.
For compact metric networks like those examined herein, the spectrum is discrete. 
Thus we look for the sets of eigenvalues $\lambda$ and eigenmodes $\mathbf{\Psi}$ satisfying
\begin{equation}
    \bigtriangleup\mathbf{\Psi} = \lambda\mathbf{\Psi}.
\end{equation}
The eigenvalues of a compact metric network such as this are unbounded from below and only a finite number are nonnegative.
The non-positive eigenvalues $\lambda=-k^2$ can be found analytically by solving 
\begin{equation}
\psi_m(x)=a_me^{ikx}+b_me^{ik(\ell_m-x)}\quad m=1,2,\dots,|\mathcal{E}|,
\end{equation}
subject to the boundary conditions described above.
In this work, however, we solve the eigensystem numerically using the QGLAB software.
An example of one of these numerically-solved eigenmodes is shown in Fig. \ref{fig:configex}.

\subsection{Hyperuniformity}
A point pattern in $d$-dimensional Euclidean space $\mathbb{R}^d$ is completely statistically characterized by the $n$-particle probability density functions $\rho_n(\mathbf{r}_1,\dots,\mathbf{r}_n)$, which are proportional to the probability of finding $n$ particles at the positions $\mathbf{r}_1,\dots,\mathbf{r}_n$ \cite{hansen_theory_1990}.
For statistically homogeneous systems, $\rho_1(\mathbf{r}_1)$ is the number density $\rho$ and $\rho_2(\mathbf{r}_1,\mathbf{r}_2)=\rho^2g_2(\mathbf{r})$, where $\mathbf{r}=\mathbf{r}_1-\mathbf{r}_2$, and $g_2(\mathbf{r})$ is the pair correlation function.
If a system is also statistically isotropic, then $g_2(\mathbf{r})=g_2(r)$ where $r=|\mathbf{r}|$.
The ensemble-averaged structure factor $S(\mathbf{k})$ is defined as 
\begin{equation}
    S(\mathbf{k}) = 1+\rho\tilde{h}(\mathbf{k}),
\end{equation}
where $\tilde{h}(\mathbf{k})$ is the Fourier transform of the total correlation function $h(\mathbf{r})=g_2(\mathbf{r})-1$.

Alternatively, one can characterize the pair statistics of a point pattern with the number variance $\sigma_N^2(R)$ associated with a spherical observation window of radius $R$.
One can either directly sample the number of points $N(R)$ in uniformly randomly positioned observation windows to numerically sample the number variance, i.e., $\sigma_N^2(R)\equiv\langle N^2(R)\rangle-\langle N(R)\rangle^2$, or compute it in terms of $g_2(\mathbf{r})$ or $S(\mathbf{k})$ \cite{torquato_local_2003}:
\begin{equation}\label{eq:nv}
    \begin{split}
        \sigma_N^2 &=\rho v_1(R)\left[1+\rho\int_{\mathbb{R}^d} h(\mathbf{r})\alpha_2(r;R)d\mathbf{r}\right], \\
        &=\rho v_1(R)\left[\frac{1}{(2\pi)^d}\int_{\mathbb{R}^d}S(\mathbf{k})\tilde{\alpha}_2(k;R)d\mathbf{k} \right],
    \end{split}
\end{equation}
where $v_1(R)$ is the volume of a sphere with radius $R$, $\alpha_2(r;R)$ is the intersection volume of two spheres of radius $R$ whose centroids are separated by a distance $r$ scaled by the volume of one such window, and $\tilde{\alpha}_2(k;R)$ is its Fourier transform.

A point pattern is hyperuniform if its angular-averaged structure factor $S(k)$, where $k\equiv|\mathbf{k}|$ is the wavenumber, tends to zero as the wavenumber tends to zero.
One can characterize the strength of hyperuniformity in a system based on the power-law scaling of $S(k)$ as $k$ tends to zero, i.e., $S(k)\sim k^\alpha$ as $k\rightarrow0$, where a hyperuniformity scaling exponent $\alpha>0$ indicates a point pattern is hyperuniform.
Equivalently, a point pattern is hyperuniform if the number variance grows more slowly than the volume of the observation window, i.e., $\sigma_N^2(R)/v_1(R)\rightarrow0$ as $R\rightarrow\infty$.
Hyperuniform point patterns can be split into three distinct classes based on the scaling behaviors of their structure factors and number variances 
\cite{torquato_local_2003}:
\begin{equation}
    \sigma_N^2(R)\sim
    \begin{cases}
        R^{d-1},\quad\alpha > 1 &\textrm{(Class I)},\\
        R^{d-1}\textrm{ln}(R),\quad\alpha = 1&\textrm{(Class II)},\\
        R^{d-\alpha},\quad0<\alpha<1&\textrm{(Class III)},
    \end{cases}
\end{equation}
where class I and III are the strongest and weakest types of hyperuniformity, respectively.

Recently, the notion of hyperuniformity was extended to treat spatially embedded network structures via the definition of a variance-based method to characterize the large-scale density fluctuations in spatially-embedded networks analogous to number variance \cite{maher_characterizing_2025}.
In particular, Maher and Newhall \cite{maher_characterizing_2025} found that networks generated from the tessellations of hyperuniform point patterns did not completely inherit the hyperuniformity of their progenitor point patterns.
However, networks derived from hyperuniform point patterns with large degrees of short-scale translational order were \textit{effectively} hyperuniform \cite{chen_binary_2018}, suggesting that they should still possess the desirable physical properties hyperuniform systems are known to possess \cite{torquato_hyperuniform_2018}.
Indeed, these desirable properties of the networks have been verified both experimentally and computationally (see e.g., Refs. \cite{torquato_hyperuniform_2018} and \cite{torquato_extraordinary_2022} and references therein for examples).
Additionally, we note that other methods have been employed to characterize the density fluctuations in networks derived from Delaunay \cite{newby_point_2025} and Voronoi \cite{newby_structural_2024} tessellations of hyperuniform and nonhyperuniform point patterns and characterize their physical and mathematical properties \cite{raj_local_2025}.

\section{Methods}\label{Sec:Methods}
In this section, we first describe the point patterns, tessellation schemes, and structural boundary conditions imposed to generate the networks examined herein.
We show representative networks of each ensemble considered herein in the Supplementary Material \cite{mmn_supp}, Figures SM$1$-SM$7$.
Then, we present the methods used to characterize the eigensystems of these metric networks with varying degrees of geometrical and topological disorder.

\subsection{Point Patterns}
We generate networks from three different point patterns in this work: two hyperuniform models and one nonhyperuniform model.
The first hyperuniform model we consider is the uniformly randomized lattice (URL) \cite{klatt_cloaking_2020}.
Here, we make the particular choice of displacing the points of the $\mathbb{Z}^2$ lattice by a random $2$-dimensional vector uniformly distributed on a scaled version of the unit cell $\mathcal{F}$ of $\mathbb{Z}^2$, i.e., $a\mathcal{F} \equiv [-a/2,a/2)^2$, where $a$ controls the perturbation strength. 
Regardless of the exact value of $a>0$, the URL is a class I hyperuniform system with $\alpha = 2$.
One can tune the degree of short-scale translational disorder in the point pattern with the specific choice of $a$, in particular, larger values of $a$ result in a greater degree of short-scale translational disorder.
In this work we use $\mathbb{Z}^2$ lattices with $N=400$ points in a square box under periodic boundary conditions as our initial conditions and values of $a$ up to 0.3.
We choose this system size because it is similar to the sizes used in the 3D-printed samples used in previous work \cite{obrero_electrical_2025} and choose this maximum value of $a$ to avoid very short edges which leads to numerical discretization issues.
Analogous considerations will apply to the rest of the point patterns described below.

Additionally, we consider a modified URL procedure where we use the $A_2$ lattice as an initial condition.
In this work, we are not interested in examining the cloaking properties of this particular scheme \cite{klatt_cloaking_2020}, so we still use a perturbation of the form $[-a/2,a/2)^2$, where $a$ is now the nearest-neighbor distance between points in the lattice.
Since we generate this point pattern in a periodic square box, which $A_2$ is not commensurate with, we choose $N = 418$ points to mitigate deformation of the $A_2$ structure in a square box \cite{uche_constraints_2004}.
Hereafter, we differentiate between these two types of URL point patterns by calling them $\mathbb{Z}^2$ URLs and $A_2$ URLs.
Figure \ref{fig:configex} depicts an eigenmode of the Laplacian on an example metric network generated from the Delaunay tessellation (described in Sec. \ref{sec:TessScheme}) of the $A_2$ URLs described here.

The second type of hyperuniform point pattern considered here is a disordered stealthy hyperuniform point pattern.
These point patterns have $S(k) = 0$ for $0 < k \leq K$ and thus are class I hyperuniform.
We generate these point patterns using the collective coordinate optimization scheme starting from totally uncorrelated initial conditions \cite{zhang_ground_2015}. 
The parameter $\chi$ is a dimensionless measure of the ratio of constrained degrees of freedom in the stealthy system (i.e., the number of wavenumbers less than $K$ constrained to be 0) to the total number of degrees of freedom (roughly $dN$, where $N$ is the number of particles in the system).
Relatively unconstrained (low-$\chi$) disordered stealthy systems will have greater translational disorder on short length scales and as $\chi$ increases within the so-called ``disordered regime'' ($0 < \chi < 1/2$ for $d=2,3$ \cite{torquato_ensemble_2015}) the degree of short length scale translational disorder decreases.
In this work we consider stealthy point patterns with $N=400$ and $\chi = 0.4, $ and 0.48.

In addition to the hyperuniform patterns above, we also examine the dense hard-disk fluid with a packing fraction of $\phi = 0.65$.
This nonhyperinform point pattern has been shown to possess an isotropic photonic bandgap, but only in finite systems \cite{froufe-perez_role_2016}.
Thus, the dense hard-disk fluid is a natural comparison for the disordered stealthy hyperuniform systems, which are known to possess isotropic photonic and phononic bandgaps \cite{florescu_designer_2009, froufe-perez_role_2016, gkantzounis_hyperuniform_2017, siedentop_stealthy_2024} that persist in the thermodynamic limit \cite{klatt_wave_2022}.
We generate these point patterns with $N = 400$ disks in a periodic square box using a standard Monte Carlo scheme.

\subsection{Tessellation Schemes}\label{sec:TessScheme}
To generate the networks examined herein, we first map point patterns under periodic boundary conditions in $\mathbb{R}^2$ to sets of polygons that tile space using the Delaunay (D), Delaunay-centroidal (C), and Gabriel (G) tessellations (described below).
Then, we produce a network from the edges that make up the boundary of each cell in the tessellation.
Each of the schemes below respects the periodic boundary conditions of the simulation box used to generate the point pattern to which we apply the tessellation scheme.
To generate the Delaunay tessellation \cite{barthelemy_spatial_2022}, we choose sets of 3 points in the point pattern and form a triangle out of them if the circumcircle of those three points does not contain another point in the pattern.
The Delaunay-centroidal tessellation \cite{florescu_designer_2009} is generated by connecting the centroid of each simplex in the Delaunay tessellation to the centroid of each neighboring simplex, i.e., those that share an edge.
Finally, the Gabriel tessellation \cite{gabriel_new_1969} is derived from the Delaunay tessellation by removing any edge if a sphere generated using that edge as its diameter contains another vertex in the tessellation.
While the Voronoi tessellation \cite{barthelemy_spatial_2022} is a natural fourth type of tessellation to examine, this scheme has a greater propensity to yield very short edges, especially in point patterns with a very small degree of short-scale translational order.
Thus, we do not use the Voronoi tessellation in this work to avoid the discretization issues mentioned in the previous subsection.

\subsection{Boundary Conditions}
We are interested in compact, i.e. nonperiodic, metric networks and thus must break the periodicity of the networks generated using the schemes above, which we do in three different ways.
1) We truncate each edge at the boundary of the simulation box. This process splits each edge that would cross the periodic boundary into two edges that terminate in degree-1 vertices. We call these \textit{clip} boundary conditions.
2) We remove any edge that crosses the periodic boundary of the square simulation box, reducing the number of edges compared to the original periodic network configuration. We call these \textit{delete} boundary conditions.
3) Starting from the networks generated from the delete boundary conditions, we additionally remove edges (if any) that terminate in a degree-1 vertex, which we call \textit{prune} boundary conditions.

\subsection{Eigensystem Computation and Characterization Methods}
With the compact networks generated via the methods described in the previous three sections in hand, we can interpret them as metric networks by assigning lengths to each edge in the network. 
For these lengths we use either the Euclidean length of each edge that one would obtain from the tessellation schemes above (referred to hereafter as \textit{standard} networks), or set each edge length to 1 yielding an \textit{equilateral} network.
While this equilateral scheme is less physically realistic, it allows us to better isolate the effects of topological disorder on the eigensystem from the effects of geometric disorder.
To ensemble average different realizations of these disordered networks, we adapt the scheme used in Ref. \cite{klatt_wave_2022} and approximate a density of states curve by
producing a histogram of the eigenvalues computed across all of the members of a particular ensemble.
In this work, since we are concerned only with the compact structures described above, we use small (i.e., less than 100 networks) ensembles of each combination of point pattern, tessellation, and network boundary conditions considered and compute the first 1500 non-positive eigenvalues and eigenmodes for each member of each ensemble.
The discretization and system sizes used herein are sufficient to prevent significant changes to the regime of the eigenvalue spectrum of interest in this work, see the Supplementary Material \cite{mmn_supp}, Section SM$2$.
We leave a study of the behaviors of the eigensystem of these metric networks in the thermodynamic limit, analogous to the one in Ref. \cite{klatt_wave_2022}, to future work.

To characterize the localization of eigenmodes in our metric networks, we compute the inverse participation ratio (IPR) following the generalization of the IPR for combinatorial networks to metric networks by Gaio et al. \cite{gaio_nanophotonic_2019}.
In particular, we use the formulation from Kravitz et al. \cite{kravitz_localized_2023},
\begin{equation}
    IPR=\frac{\sum_m\int_0^{\ell_m}\psi^4_m(x)\textrm{d}x}{\left(\sum_m\int_0^{\ell_m}\psi^2_m(x)\textrm{d}x\right)^2},
\end{equation}
where $\psi(x)_m$ is the portion of a particular eigenmode $\mathbf{\Psi}$ on edge $m$ and $m$ indexes over all edges in the network.
Values of the IPR closer to 1 indicate strong eigenmode localization, while values closer to 0 indicate delocalization.

\section{Results and Discussion} \label{Sec:Results}

In this section, we characterize the eigensystem of the metric networks described in Sec. \ref{Sec:Methods}.
Specifically, we examine the changes in the structure of the density of eigenvalues as we change the point patterns, tessellation schemes, and network boundary conditions used to generate compact metric networks.
We then determine what structural characteristics of these metric networks result in deep troughs in the density of eigenvalues.
Here, we focus on troughs in the eigenvalue density and thus overlay the eigenvalue densities such that those with the widest troughs are on top (where possible), and leave a more detailed discussion of other features of eigenvalue density to future work.
We present the individual (i.e., not overlaid) eigenvalue densities in the Supplementary Material \cite{mmn_supp}, Section SM$3$.

\subsection{Comparison of Different Tessellations and Network Boundary Conditions}

\begin{figure}[!b]
    \centering
    \subfigure[]{\includegraphics[width=0.45\linewidth]{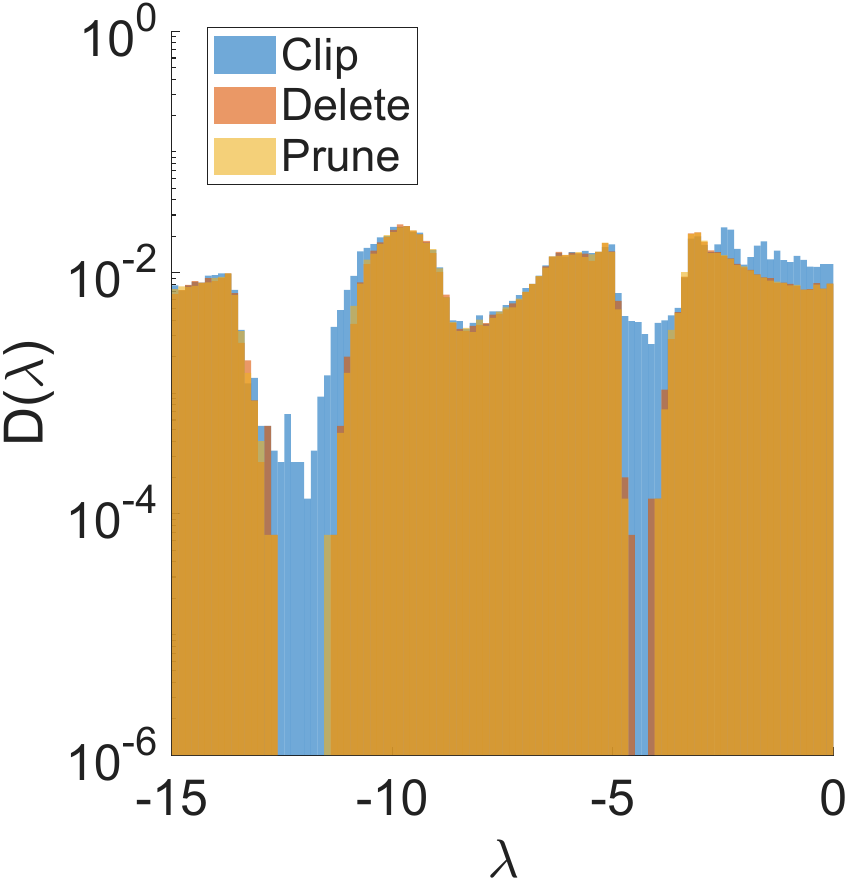}}
    \subfigure[]{\includegraphics[width=0.45\linewidth]{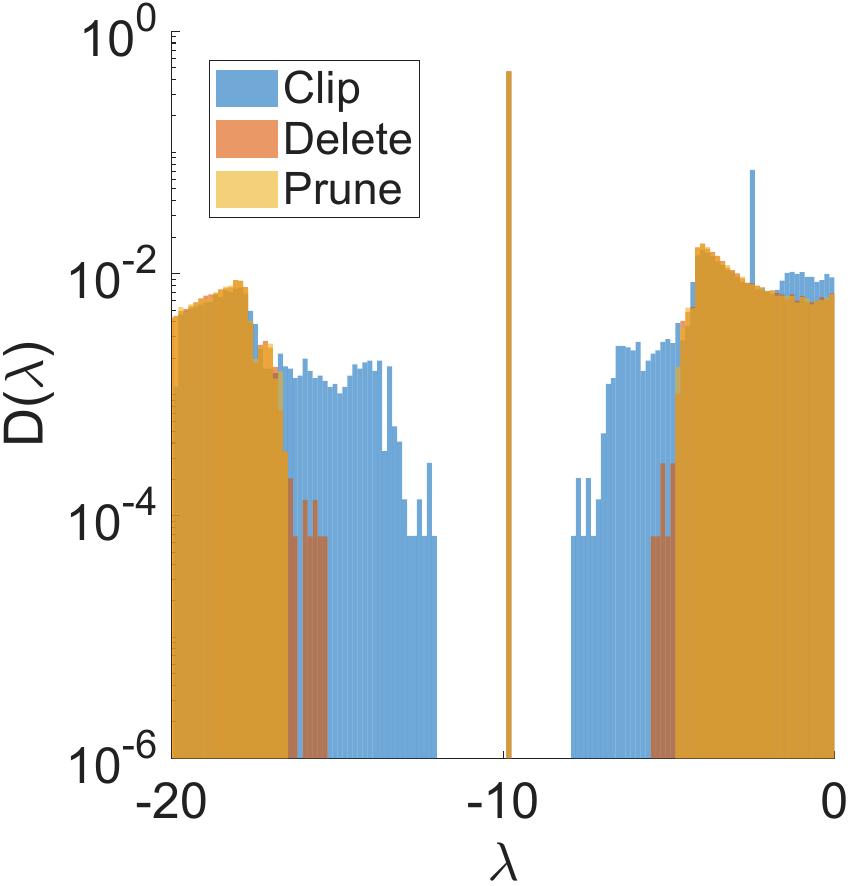}}
    \caption{Density $D(\lambda)$ of eigenvalues $\lambda$ for (a) standard and (b) equilateral metric networks generated from a Delaunay tessellation of a $\mathbb{Z}^2$ URL with $a=0.1$ with either clip, delete, or prune boundary conditions applied.}
    \label{fig:Boundary}
\end{figure}

First, given our sets of tessellation and network boundary types, we aim to determine the combination that results in the deepest and widest troughs in the eigenvalue densities.
In Fig. \ref{fig:Boundary}, we compare the clip, delete, and prune boundary conditions for (a) standard and (b) equilateral metric networks generated from the Delaunay tessellation of a $\mathbb{Z}^2$ URL with $a = 0.1$.
In this region of the eigenvalue spectrum, we observe that the spectrum is split into three parts where there is a central ``island'' of states with two troughs on either side.
The depth and width of these troughs change with different boundary conditions.
For the standard metric networks, we find that the delete and prune boundary conditions result in similar eigenvalue densities, with a central ``island'' with very deep troughs on either side at $-4.5\lesssim\lambda\lesssim-4.2$ and $-12.5\lesssim\lambda\lesssim-11.5$.
By contrast, the clip boundary conditions result in much shallower troughs that are located at approximately the same eigenvalues.
For the equilateral networks, notice that the central island of eigenvalues is now a single sharp peak at $\lambda\approx10$.
The troughs for the clip boundary conditions are much deeper, but not as wide as those for the delete boundary conditions, which themselves are not as wide as those for the prune boundary conditions.
Given these observations, we use the prune boundary conditions in the remainder of our computations.

\begin{figure}[!b]
    \centering
    \subfigure[]{\includegraphics[width=0.45\linewidth]{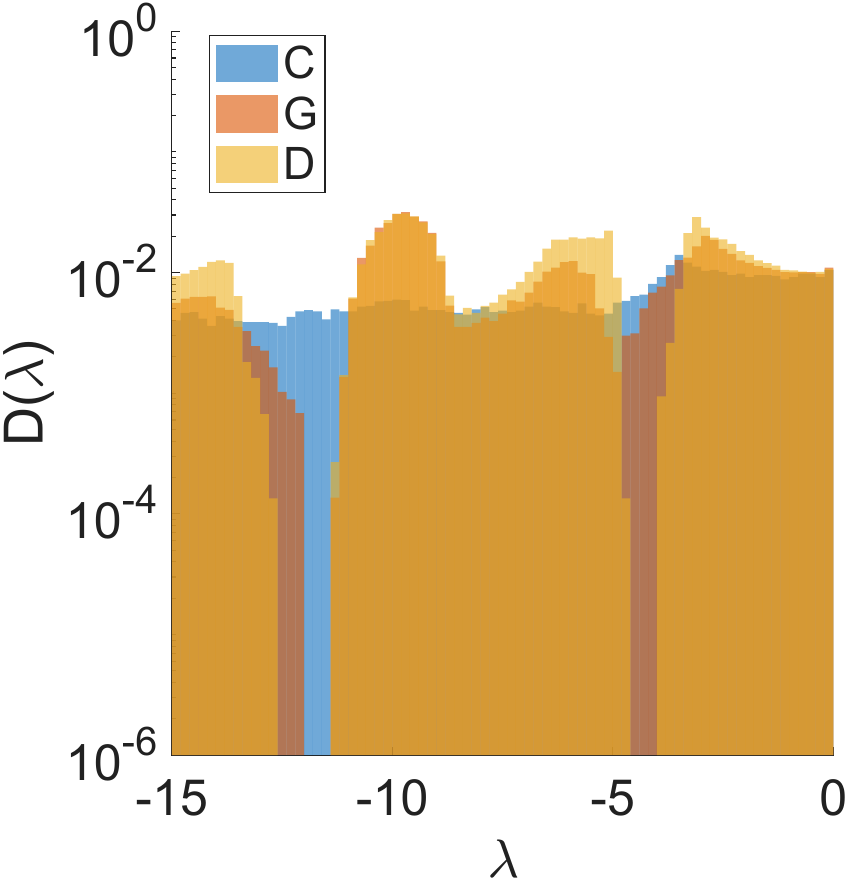}}
    \subfigure[]{\includegraphics[width=0.45\linewidth]{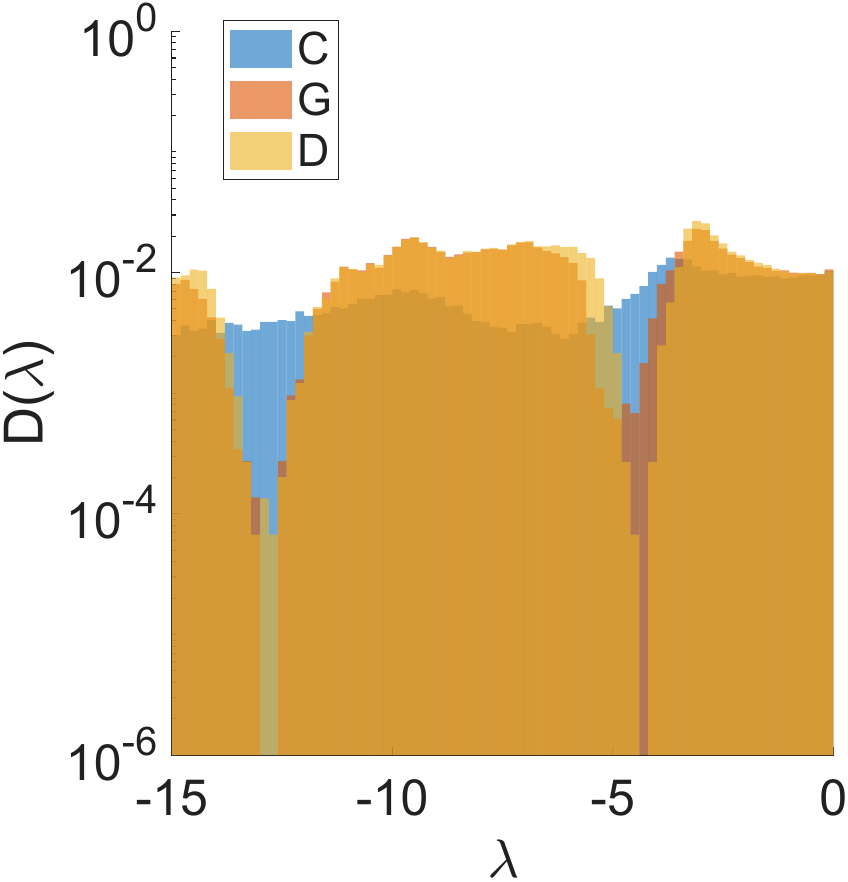}}    
    \caption{Density $D(\lambda)$ of eigenvalues $\lambda$ for metric networks generated from a Delaunay-centroidal (C), Gabriel (G), or Delaunay (D) tessellation of (a) a $\mathbb{Z}^2$ URL with $a=0.1$ and (b) a disordered stealthy hyperuniform point pattern with $\chi = 0.48$, both with prune boundary conditions.}
    \label{fig:Tess}
\end{figure}

In Fig. \ref{fig:Tess}(a), we show the eigenvalue densities for Delaunay (D), Gabriel (G), and Delaunay-centroidal (C) tessellations of a $\mathbb{Z}^2$ URL point pattern with $a = 0.1$.
We find that the C tessellation yields a relatively flat eigenvalue density, the G tessellation has one deep trough at $\lambda\approx-11.7$ and a shallow trough at $\lambda\approx-5$, and the behavior of the D tessellation is as described above.
In Fig. \ref{fig:Tess}(b), we see that the C tessellation of a disordered stealthy hyperuniform point pattern with $\chi = 0.48$ results again in a relatively flat eigenvalue density, while the D and G tessellations have two troughs.
In addition, while the D tessellation has a deep narrow trough at $\lambda\approx-4$ and a shallow trough at $\lambda\approx-13$, the G tessellation results in the opposite behavior.
Interestingly, the spectral behavior of the C tessellation does not agree with the intuition one would have from the heterogeneous materials literature, which has used the C tessellation of stealthy hyperuniform point patterns to produce materials with photonic bandgaps \cite{florescu_designer_2009, froufe-perez_role_2016, siedentop_stealthy_2024}.
Along similar lines, we find that the $\mathbb{Z}^2$ has deeper and wider troughs than the disordered stealthy hyperuniform patterns, which also runs counter to the intuition from the hyperuniform photonic materials literature.
These findings suggest that the lessons learned from the materials literature may need to be modified to apply to these metric network structures.
Due to the presence of two deep troughs in eigenvalue densities of the metric networks generated from D tessellations of the URL, we focus on these tessellations in the next subsection.

\subsection{Effects of Geometric Disorder}

\begin{figure}[!t]
    \centering
    \subfigure[]{\includegraphics[width=0.45\linewidth]{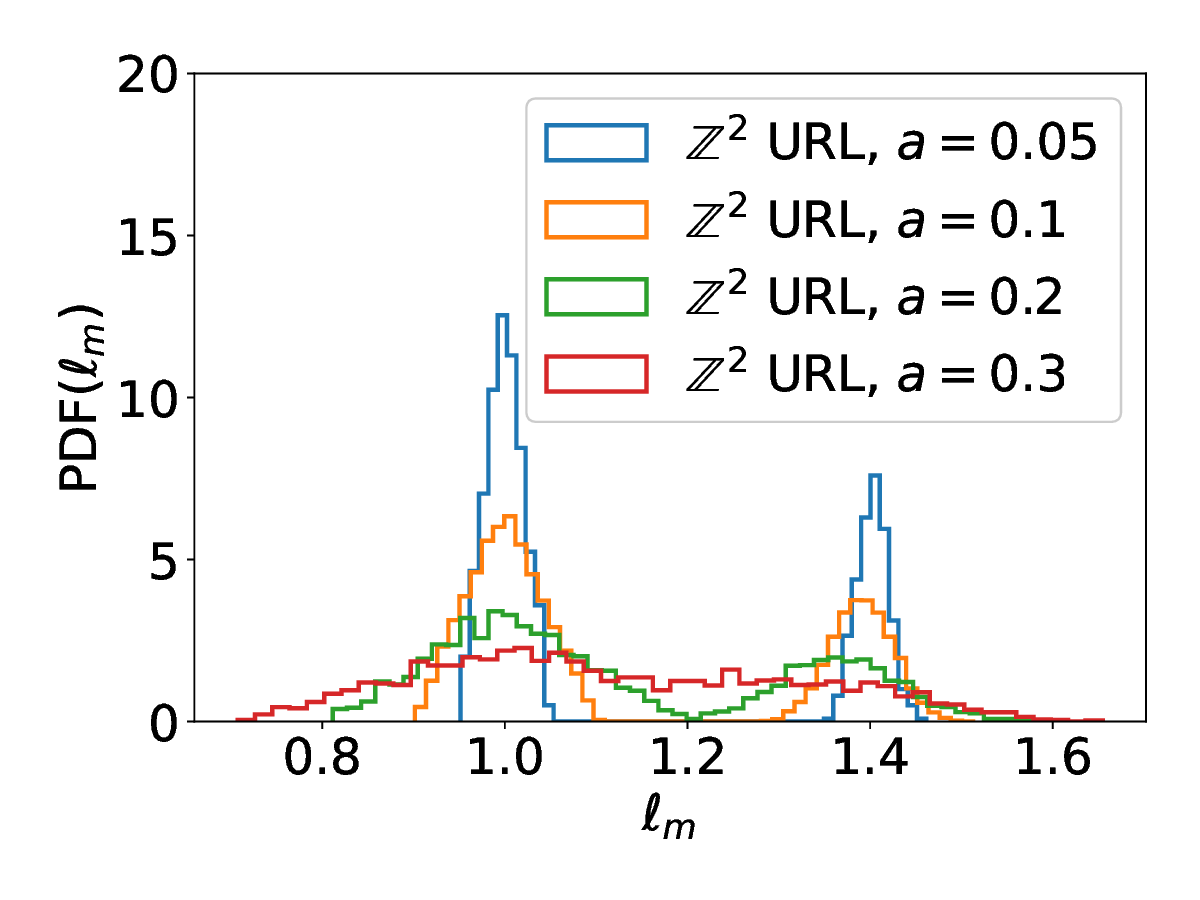}}
    \subfigure[]{\includegraphics[width=0.45\linewidth]{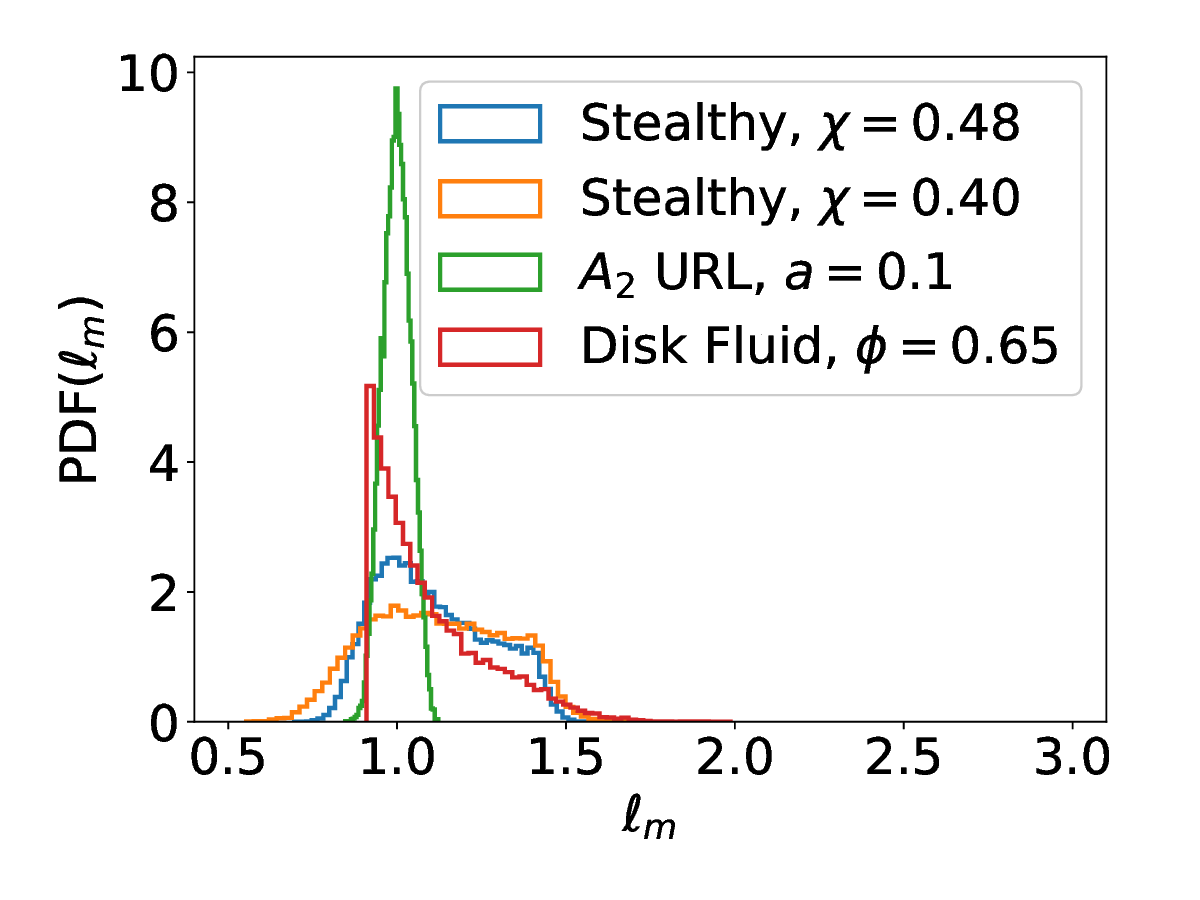}}
    \caption{Probability density of edge lengths $\ell_m$ in D networks generated from (a) $\mathbb{Z}^2$ URLs with $a = 0.05,0.1,0.2,$ and $0.3$ and (b) disordered stealthy hyperuniform point patterns with $\chi = 0.48$ and $0.40$, $A_2$ URLs with $a = 0.1$ and hard-disk fluids with $\phi = 0.65$.}
    \label{fig:ELD}
\end{figure}

To probe how geometric disorder in the metric networks affects the eigenvalue density, we generated metric networks from a variety of different point patterns using the D tessellation and prune boundary conditions.
These different point patterns yield different distributions of edge lengths, which we display in Fig. \ref{fig:ELD}.
For URL point patterns, the distribution of edge lengths broadens as $a$ increases, which is consistent with the increase in short-scale translational disorder in the point pattern that occurs when $a$ is increased.
We only show the edge-length distributions explicitly for the $\mathbb{Z}^2$ URLs for the sake of brevity.
We also compare the edge-length distributions for disordered stealthy hyperuniform point patterns with $\chi = 0.48$ and $0.4$, hard-disk fluids with $\phi = 0.65$, and $A_2$ URLs with $a = 0.1$, which are all monomodal.
The edge-length distribution for the disk fluid is bounded from below due to the nonoverlap constraint of the hard disks, and has a maximum edge length larger than the other three point patterns.
The $A_2$ URL with $a = 0.1$ has a sharp distribution centered on the $A_2$ lattice spacing (1, in this case).
The disordered stealthy hyperuniform point patterns have a broader edge-length distributions than the $A_2$ URL, and are less skewed than the hard-disk edge-length distribution.
Additionally, the width of the distribution for the disordered stealthy hyperuniform point patterns decreases with increasing $\chi$, which is consistent with the increase in short-scale translational order in the point patterns as $\chi$ increases.

\begin{figure}[!t]
    \centering
    \subfigure[]{\includegraphics[width=0.32\linewidth]{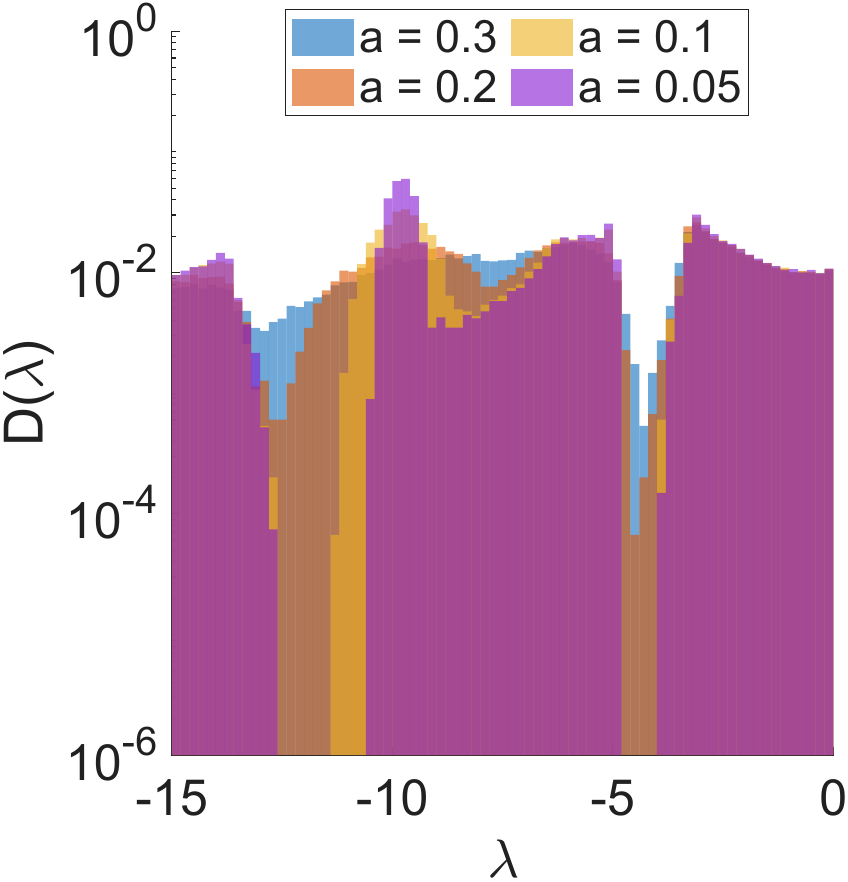}}
    \subfigure[]{\includegraphics[width=0.32\linewidth]{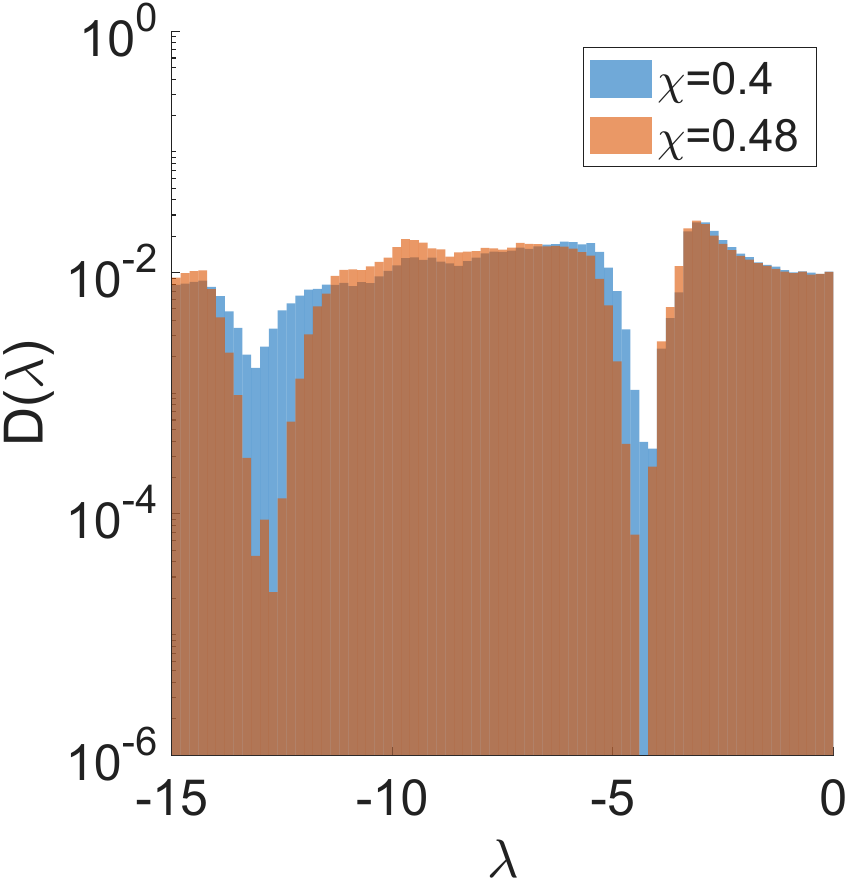}}
    \subfigure[]{\includegraphics[width=0.32\linewidth]{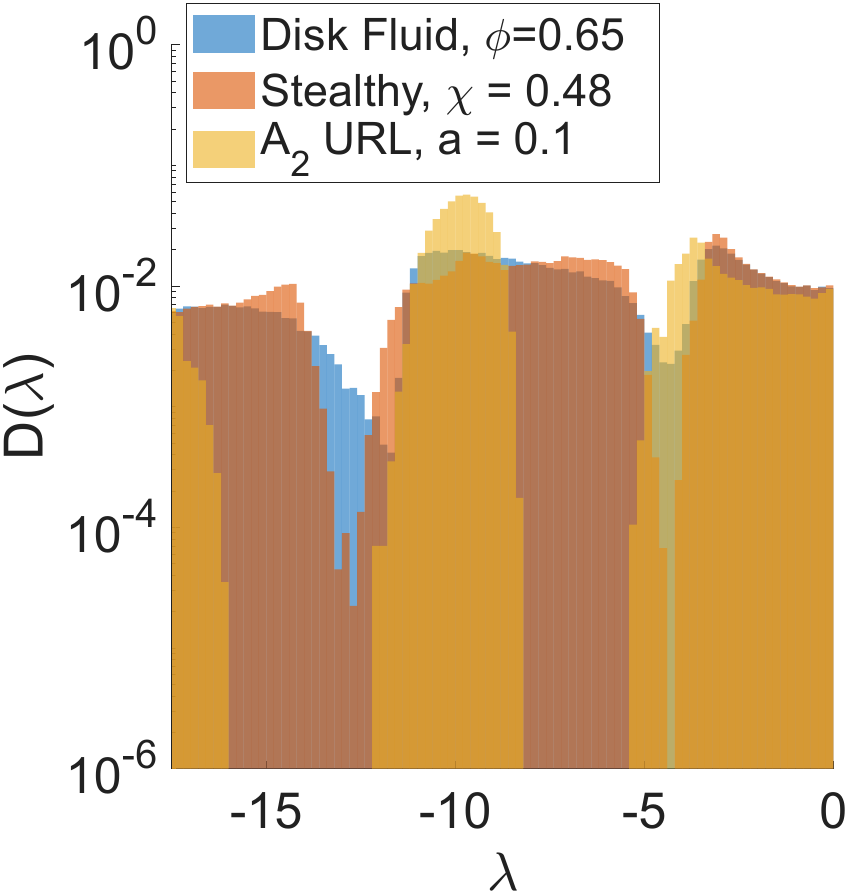}}
    \caption{Density $D(\lambda)$ of eigenvalues $\lambda$ for metric networks generated from Delaunay tessellations of (a) $\mathbb{Z}^2$ URLs with $a = 0.05,0.1, 0.2$ and $0.3$ (b) disordered stealthy hyperuniform point patterns with $\chi = 0.48$ and $0.40$ and (c) an $A_2$ URL with $a = 0.1$, a disordered stealthy hyperuniform point pattern with $\chi = 0.48$, or a dark disk fluid point pattern with $\phi = 0.65$ with prune boundary conditions.}
    \label{fig:Pattern}
\end{figure}

To examine how these edge-length distributions affect the eigenvalue distributions, we first compare these distributions for metric networks derived from the D tessellations of $\mathbb{Z}^2$ URLs with values of $a$ between 0.05 and 0.3 in Fig. \ref{fig:Pattern}(a).
Clearly, one can observe that as $a$ increases, which broadens the edge-length distribution, the troughs in the corresponding eigenvalue densities become more shallow.
We carry out an analogous experiment on disordered stealthy hyperuniform point patterns with $\chi = 0.48$ and 0.4, whose eigenvalue distributions are shown in Fig. \ref{fig:Pattern}(b). 
Comparing the edge-length distributions for the metric networks generated from the D tessellations of disordered stealthy hyperuniform point patterns in Fig. \ref{fig:ELD}(b) to the eigenvalue distributions in Fig. \ref{fig:Pattern}(b), we observe the same behavior: as $\chi$ decreases, which broadens the edge-length distribution, the troughs in Fig. \ref{fig:Pattern}(b) become more shallow.
Moreover, in Fig. \ref{fig:Pattern} (c), we show that the relationship between broader edge-length distributions and shallower troughs holds across networks derived from different kinds of initial point patterns.
Overall, we observe that a broader edge-length distribution results in a broader central ``island'' in the eigenvalue densities, causing the troughs on either side of the island to become more shallow.
These findings suggest that, in metric networks derived from tessellations, it is important to have narrow edge-length distributions to obtain deep troughs in the eigenvalue spectrum.
Notably, the $\mathbb{Z}^2$ URL results also suggest that it is not necessary to have a narrow, monomodal edge-length distribution and one can achieve deep troughs with bimodal edge-length distributions, so long as the distribution around each mode is narrow.

\subsection{Effects of Topological Disorder}
Now, we aim to determine how topological changes in our metric networks can affect the eigenvalue density.
Since it has been shown that the girth, or length of the shortest cycle in a metric network, can affect its scattering behavior \cite{gnutzmann_topological_2013}, we examine how differences in the numbers of edges that surround the faces in our planar tessellations affect the eigenvalue density.
Recall from Fig. \ref{fig:Tess} that the D tessellation tends to have the deepest troughs and is composed of strictly triangles, and the G tessellation, which is generated by removing certain edges from the D tessellation, tends to have the second-deepest troughs.
Therefore, we are particularly interested in the effects of triangular faces on the eigenvalues of these 2D metric networks.

\begin{figure}[!t]
    \centering
    \subfigure[]{\includegraphics[width=0.375\linewidth]{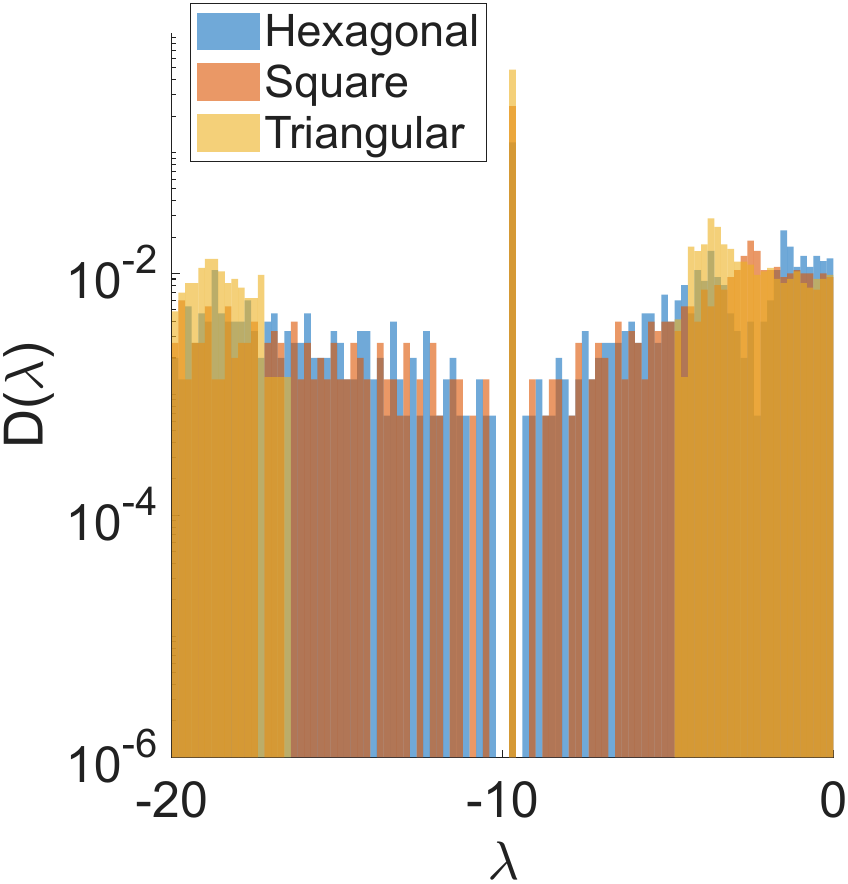}}
    \subfigure[]{\includegraphics[width=0.375\linewidth]{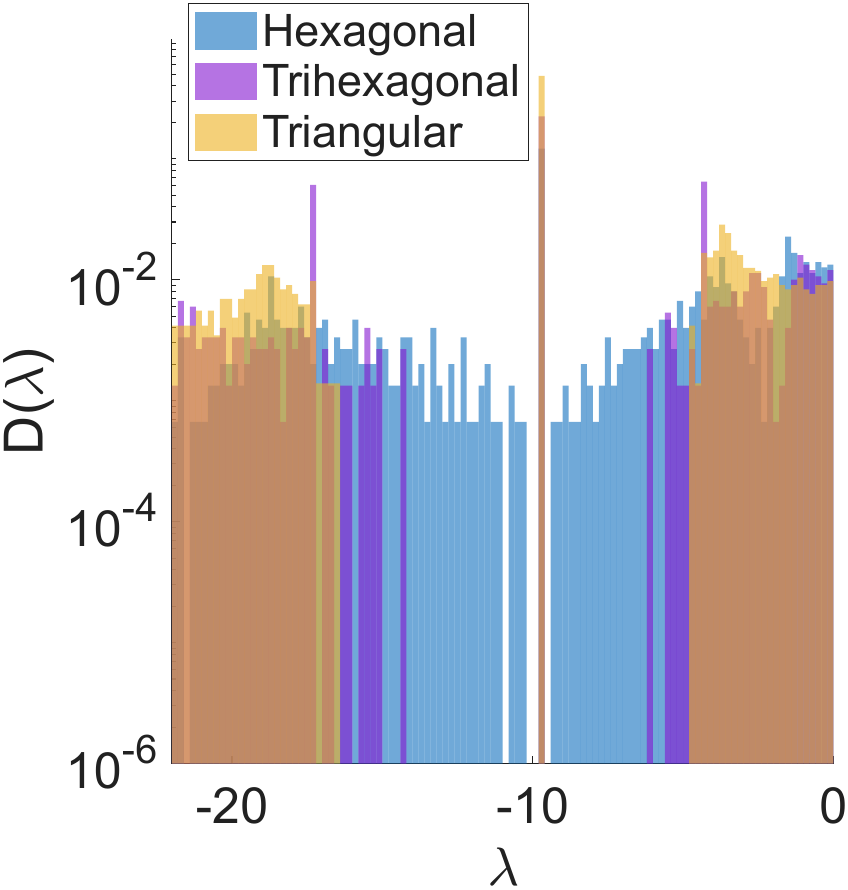}}
    \caption{Density $D(\lambda)$ of eigenvalues $\lambda$ for metric networks generated from finite portions of the (a) hexagonal, square, and triangular tilings and (b) hexagonal, trihexagonal (Kagome), and triangular tilings.}
    \label{fig:Tilings}
\end{figure}

To this end, we first compare the spectra of finite portions of the triangular tiling, the hexagonal tiling, the square tiling, and the trihexagonal (Kagome) tiling.
In Fig. \ref{fig:Tilings}(a), we compare the spectra for finite portions of the three regular tilings of $\mathbb{R}^2$ using only a single shape.
We find that, of these three, the triangular tiling has significantly wider troughs than the square and hexagonal tilings.
In Fig. \ref{fig:Tilings}(b) we compare the trihexagonal tiling, which contains both hexagons and triangles, to the triangular and hexagonal tilings, and find that the width of the troughs for the trihexagonal tilings is greater than that of the hexagonal tiling and smaller than that of the triangular tilings.
This finding supports our hypothesis that triangular faces are important to the formation of deep troughs in the eigenvalue densities.

\begin{figure}[!t]
    \centering
    \subfigure[]{\includegraphics[width=0.375\linewidth]{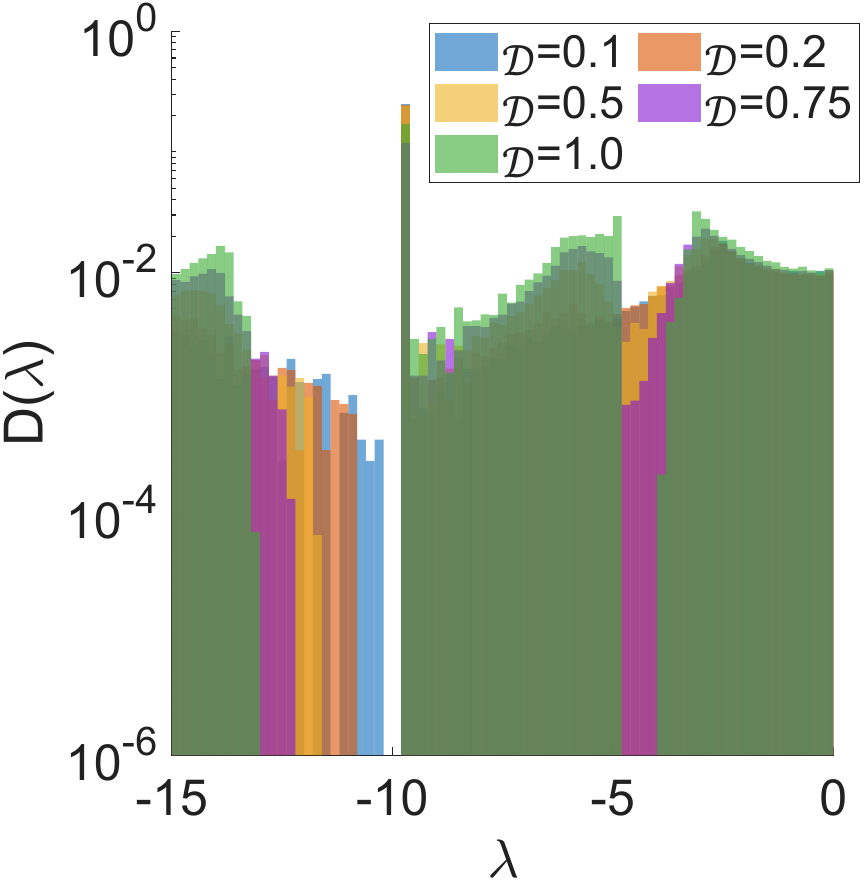}}
    \subfigure[]{\includegraphics[width=0.375\linewidth]{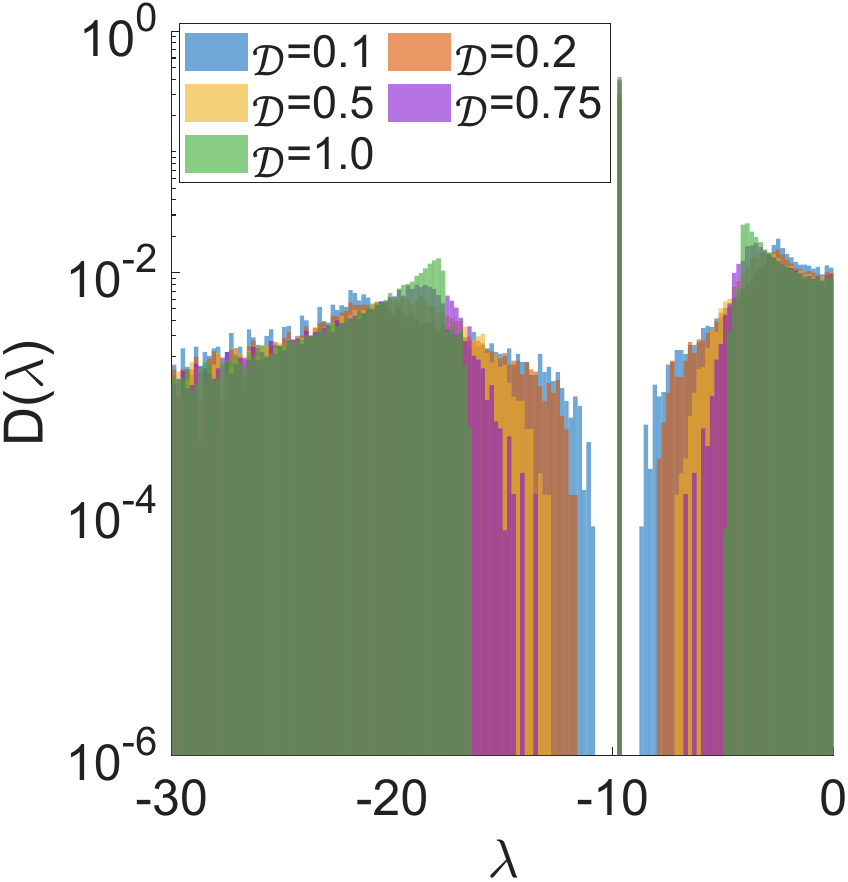}}
    \caption{Density $D(\lambda)$ of eigenvalues $\lambda$ for (a) standard and (b) equilateral metric networks generated from finite portions of the square tiling with a fraction $\mathcal{D}$ of the squares containing a randomly oriented diagonal edge.}
    \label{fig:TriDef}
\end{figure}

To examine the effect of the triangular face concentration on the eigenvalue distribution in disordered networks, we started with a finite portion of a square lattice and then added a single randomly oriented diagonal edge to a fraction $\mathcal{D}$ of the square cells in the lattice.
This procedure allows us to, in effect, add triangular defects to the square tiling.
We generated an ensemble of 10 configurations using the above procedure for each defect concentration we consider here.
In Fig. \ref{fig:TriDef} we examine the effect of these triangular defects on the eigenvalue density.
In both the standard and equilateral metric network cases, we observe that the trough widths increase monotonically with $\mathcal{D}$, which further supports the hypothesis stated above.

\subsection{Eigenmode Localization}

\begin{figure}[!b]
    \centering
    \subfigure[]{\includegraphics[width=0.4\linewidth]{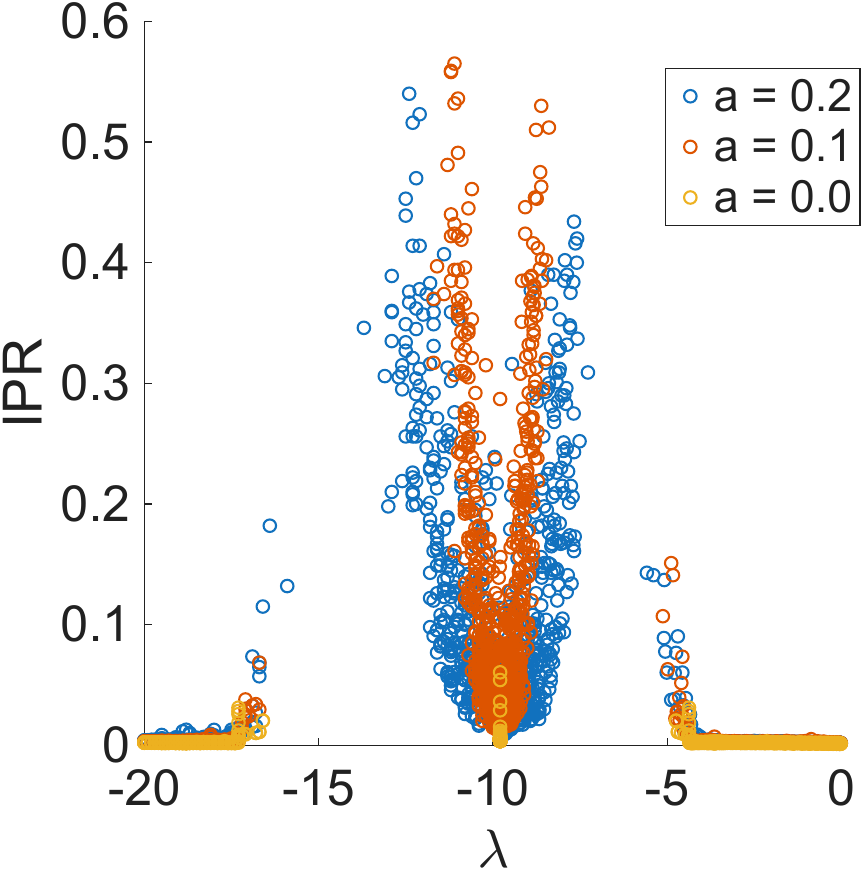}}
    \subfigure[]{\includegraphics[width=0.4\linewidth]{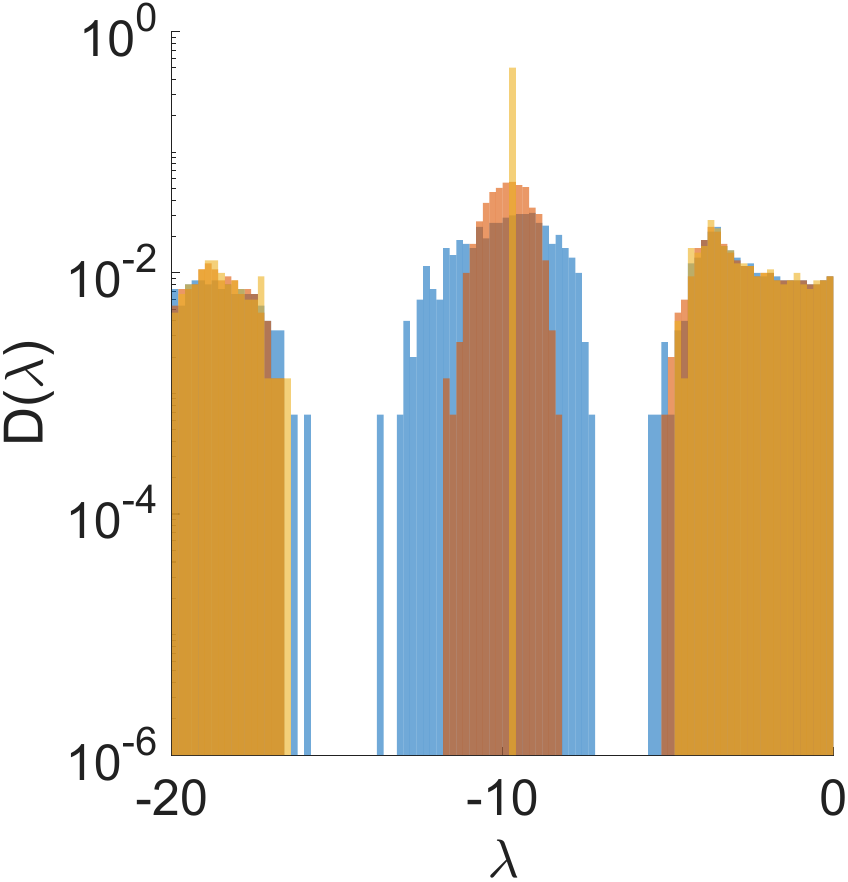}}
    \caption{The (a) Inverse participation ratio IPR as a function of eigenvalue $\lambda$ for metric networks derived from D tessellations of $A_2$ URLs with $a = 0.0,0.1$ and 0.2 and (b) the corresponding density $D(\lambda)$ of eigenvalues $\lambda$. Note that these data are from single configurations.}
    \label{fig:IPR}
\end{figure}

To characterize the eigenmodes of our metric networks, we examine their localization using the inverse participation ratio (IPR). 
In Fig. \ref{fig:IPR}, we show (a) the IPR and (b) corresponding eigenvalue densities for $A_2$ URLs with $a = 0.0,0.1$ and 0.2.
We observe two notable behaviors.
First, none of the eigenmodes in the ordered $a = 0.0$ network are strongly localized, even on the central island.
These central island states are slightly more strongly localized than those not on the island, but still have IPR values far from 1.
Second, the strongest localization occurs on the edges of the troughs in the eigenvalue density, particularly on the edges of the central island in the eigenvalue density.
The degree of localization drops off rapidly as $\lambda$ deviates from the trough edges.

\section{Conclusions and Outlook}\label{Sec:Conclusions}

In this work, we generated compact metric networks with a variety of different geometrical and topological characteristics, and examined how these characteristics affected the eigensystem of these networks.
To generate these networks, we first generated Delaunay, Gabriel, and Delaunay-centroidal tessellations under periodic boundary conditions tessellations of hyperuniform and nonhyperuniform point patterns with varying degrees of short-scale translational disorder.
Then, to produce compact networks, we broke the periodic boundary conditions to produce networks with either clip, delete, or prune boundary conditions.
To treat these tessellations as metric networks, we assigned lengths to each edge by either using the Euclidean length of each edge, or simply setting each edge length to 1.
With these various metric networks in hand, we then numerically computed the eigenvalue spectrum and corresponding eigenmodes of the Laplacian and characterized the eigenvalue density and eigenmodes.

For the point-pattern-generated networks described above, we found that the Delaunay tessellation combined with prune boundary conditions resulted in the deepest troughs in the eigenvalue density.
We are particularly interested in these deep troughs due to the importance of band and spectral gaps across disciplines \cite{kuchment_spectral_1999,kuchment_spectra_2007,lawrie_quantum_2022,yoshitomi_band_1998}.
We then generated Delaunay tessellations of hyperuniform URL, disordered stealthy hyperuniform, and nonhyperuniform dense hard-disk fluid point patterns and found that the troughs in the eigenvalue density are wider if the edge-length distributions are narrower.
These narrow edge-length distributions correspond to D tessellations generated from point patterns that have a greater degree of short-scale translational order.
We also found that metric networks with bimodal edge-length distributions can also have deep troughs in their eigenvalue densities, so long as the distribution around each mode is narrow.
We note here that the deepest and widest troughs occurring for the Delaunay tessellation is not what one would expect given the intuition from network bandgap materials \cite{florescu_designer_2009, froufe-perez_role_2016, klatt_wave_2022}.
We expect that this deviation from the materials literature is because we are now studying behaviors on a complex of 1D spaces, as opposed to in $\mathbb{R}^d$.
One can interpret the materials-based bandgap problem as a ``dual'' problem to the problem considered in this work, in the sense that we are concerned with a wave traveling \textit{through} a network structure as opposed to traveling through a $d$-dimensional Euclidean space and interacting with the network structure. This interpretation may explain why we observed the deepest troughs in the eigenvalue densities of metric networks generated from the Delaunay tessellation, while wide photonic bandgaps in disordered media tend to be observed in network materials generated from the Delaunay-centroidal tessellation \cite{florescu_designer_2009, klatt_wave_2022, froufe-perez_role_2016}.
As such, we must develop a different set of design principles to capture the expected behaviors of hyperuniform systems in metric networks.

From our study of the three tessellation schemes, we noticed that an increased fraction of triangular faces in our networks resulted in wider and deeper troughs.
To probe this behavior, we computed the eigensystems for finite tilings with different topologies: the triangular, hexagonal, square, and trihexagonal (Kagome) tilings.
We found that the triangular tiling had the widest troughs, followed by the trihexagonal tiling (which has some triangles), and the hexagonal and square tilings (which have no triangles) had the narrowest troughs.
To determine if this trend holds in disordered systems, we added randomly oriented triangular defects to the square tiling, and found that the trough widths in the eigenvalue spectra increased monotonically with the concentration of triangular defects.
These findings support the hypothesis that triangular cells in metric networks increase the depth and width of troughs in the eigenvalue density.

With this work, we have introduced a new family of compact disordered metric networks generated by the tessellations of hyperuniform point patterns.
Our characterization of the spectra of these physically-relevant networks has revealed methods by which one can tune the position and size of gaps in their eigenvalue spectra.
Moreover, because of the relationship between the solutions of linear PDEs and the eigenvalues and eigenmodes of the differential Laplacian, our findings can be used to inform the design of metric network-based systems with, e.g., prescribed heat and wave transport behaviors.
To build on our findings and increase the number of areas to which they can apply, several additional studies can be undertaken.
For example, an important area of future study is examining if these troughs observed in the eigenvalue densities persist in the thermodynamic limit, following the procedure laid out in Ref. \cite{klatt_wave_2022}.
In addition, expanding the design space to consider tessellations of point patterns in $\mathbb{R}^3$ will be important to determine how the lessons learned herein carry over into different space dimensions.

\section*{Acknowledgments}
The authors would like to thank Mason A. Porter and Lior Alon for helpful discussions. 

\section*{Data Availability}
The QGLAB software used to compute the eigensystems is available at https://github.com/manroygood/Quantum-Graphs/. The scripts used to generate the network configurations and run the QGLAB computations are openly available \footnote{(Will be available upon publication.)}.
The hard disk fluid and disordered stealthy hyperuniform configurations are available at \cite{GitHub_QGLAB_script}.

\bibliographystyle{siamplain}
\bibliography{refs_addnl}



\section*{Supplementary material}

# SUPPLEMENTARY MATERIALS: Probing the influence of topological and geometric disorder on the spectrum of the differential Laplacian operator on networks[*]

Charles Emmett Maher[†], Jeremy L. Marzuola[†], and Katherine A. Newhall[†]

**SM1. Representative Networks.** Here, we show representative members of the ensembles of disordered networks examined in this work. The point patterns, tessellation schemes, and network structure boundary conditions imposed are described in Sec. III of the main text with the exception of the square tiling with triangular defects, which is described in Sec. IVC.

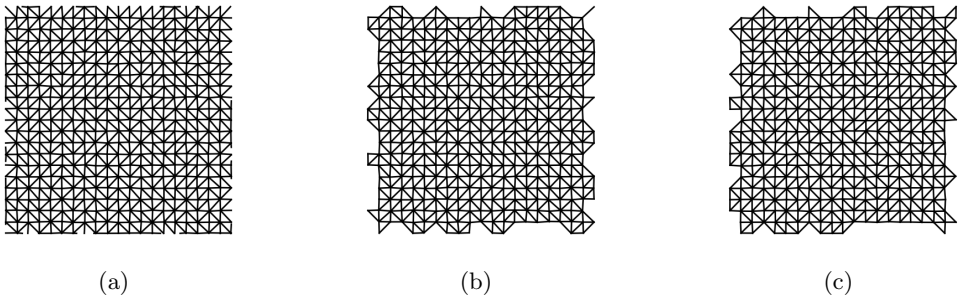

(a)　　　　　　　(b)　　　　　　　(c)

**Figure SM1.** *Representative examples of networks generated from the Delaunay tessellation of $\mathbb{Z}^2$ URLs with $a = 0.1$ and the three different types of network structure boundary conditions: (a) clip, (b) delete, and (c) prune.*

[†]Department of Mathematics, University of North Carolina at Chapel Hill, Chapel Hill, NC, USA (cemaher@email.unc.edu, marzuola@email.unc.edu, knewhall@email.unc.edu).

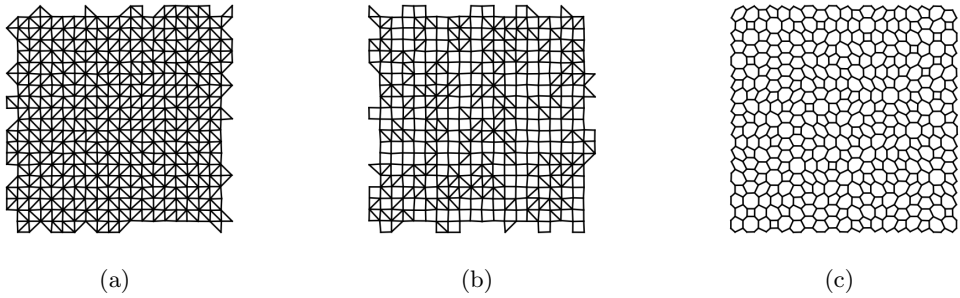

(a) (b) (c)

**Figure SM2.** *Representative examples of networks generated from the (a) Delaunay, (b) Gabriel, and (c) Delaunay-centroidal tessellations of $\mathbb{Z}^2$ URLs with $a = 0.1$ and prune network structure boundary conditions.*

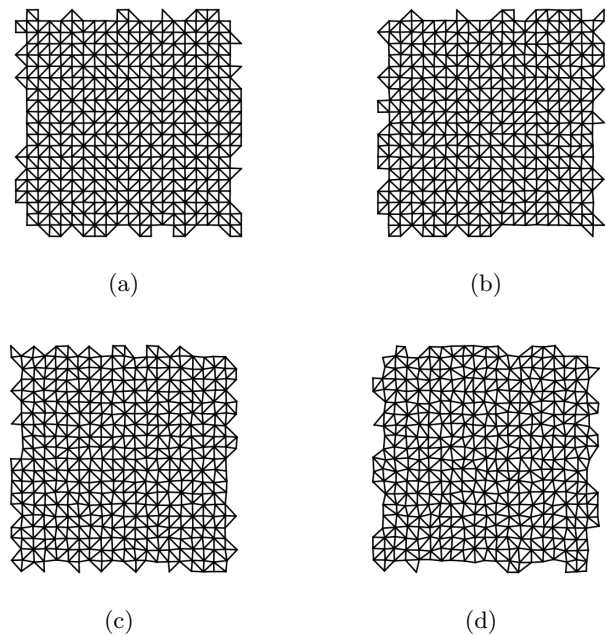

(a) (b)

(c) (d)

**Figure SM3.** *Representative examples of networks generated from the Delaunay tessellation of $\mathbb{Z}^2$ URLs with $a =$ (a) 0.05, (b) 0.1, (c) 0.2, and (d) 0.3 and prune network structure boundary conditions.*

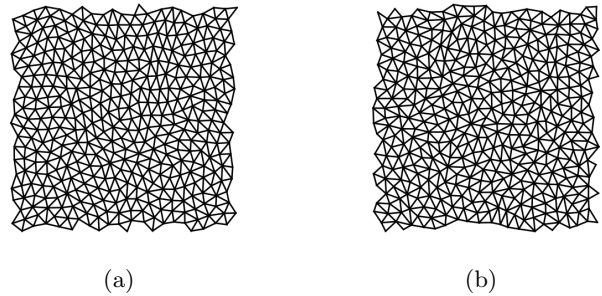

(a)          (b)

**Figure SM4.** *Representative examples of networks generated from the Delaunay tessellation of disordered stealthy hyperuniform point patterns with $\chi =$ (a) 0.48 and (b) 0.4 and prune network structure boundary conditions.*

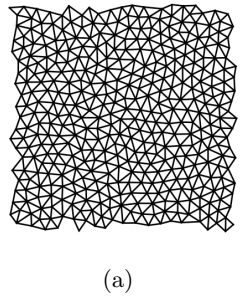

(a)

**Figure SM5.** *Representative example of a network generated from the Delaunay tessellation of an equilibrium hard disk fluid with with $\phi = 0.65$ and prune network structure boundary conditions.*

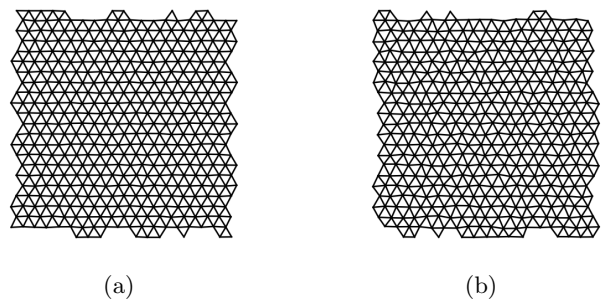

(a)          (b)

**Figure SM6.** *Representative examples of networks generated from the Delaunay tessellation of $A_2$ URLs with $a =$ (a) 0.1 and (b) 0.2 and prune network structure boundary conditions.*

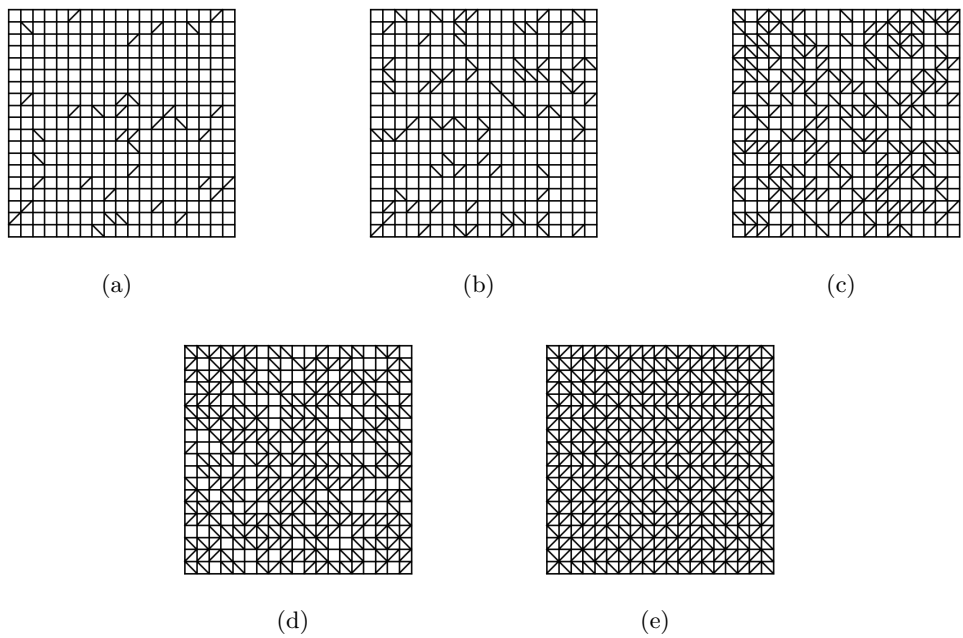

**Figure SM7.** *Representative examples of of the square tiling networks with triangular defect concentrations of (a) 0.1, (b) 0.2, (c) 0.5, (d) 0.75, and (e) 1.0*

**SM2. Effect of Discretization and System Size on the Eigenvalue Spectrum.** To use the QGLAB software to numerically compute the eigenvalue spectrum and the corresponding eigenfunctions, we need to discretize our network structure. In the main text, we use a uniform discretization resolution of one point every 0.1 length units (cf. Fig. 4 of the main text, which shows the edge-length distributions of our networks). Thus, each edge has at least 5 discretization points, with distributions centered at 10-14 points. To ensure that this discretization scheme is sufficient to avoid numerical effects in the eigenvalue spectrum, we compare the scheme used in the main text to: 1) doubling the resolution to a discretization point every 0.05 length units and 2) a Chebyshev discretization scheme with 10 points on each edge. We apply each scheme to a $\mathbb{Z}^2$ URL with $N = 400$ points and $a = 0.1$. Additionally, we ensure that we are not incurring significant finite-size effects with the $N = 400$ by testing a system roughly twice as large with $N = 784$ points. The corresponding eigenvalue densities can be found in Fig. 8. Each density is for an ensemble of 10 configurations.

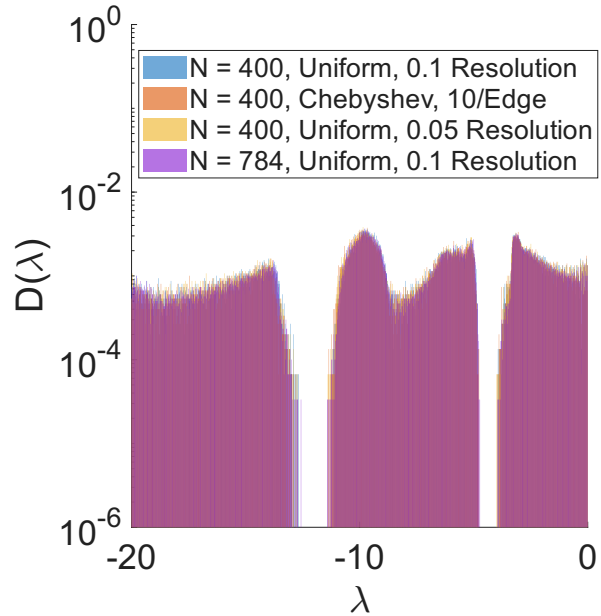

**Figure SM8.** *Density $D(\lambda)$ of eigenvalues $\lambda$ for standard metric networks generated from Delaunay tessellation of a $\mathbb{Z}^2$ URL with $a = 0.1$ and (blue) $N = 400$ points with a uniform discretization of 1 grid point every 0.1 length units, (orange) $N = 400$ points with a Chebyshev discretization of 10 grid points per edge, (yellow) $N = 400$ points with a uniform discretization of 1 grid point every 0.05 length units, and (purple) $N = 784$ points with a uniform discretization of 1 grid point every 0.1 length units.*

In Fig. 8 we use a much narrower bin width than in the main text (1/10th the size) to more clearly show the locations of the edges of the troughs in the eigenvalue densities. We observe that the overall shape of the eigenvalue density is not significantly impacted by the improvement of discretization of increase in system size. This observation suggests that the choices made for discretization and system size in the main text allow us to accurately capture the spectral behavior of the new class of metric network examined in this work.

**SM3. Individual Eigenvalue Density Distributions.** Here, we present the eigenvalue densities $D(\lambda)$ presented in the main text in individual figures so features other than the gaps (or lack thereof) in the densities are more visually apparent. We make no further comment on any additional features however, as this is outside the scope of the present work.

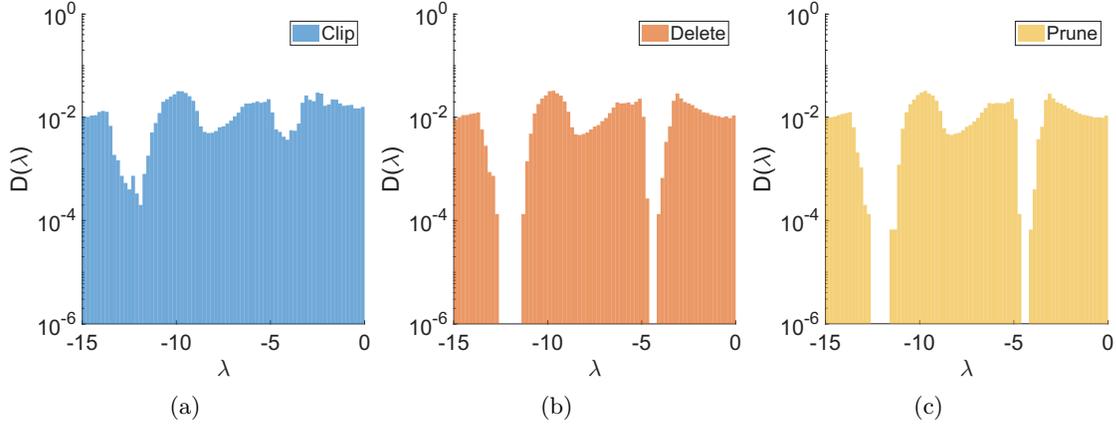

**Figure SM9.** *Density $D(\lambda)$ of eigenvalues $\lambda$ for standard metric networks generated from a Delaunay tessellation of a $\mathbb{Z}^2$ URL with $a = 0.1$ and (a) clip, (b) delete, and (c) prune network structure boundary conditions. The color of each figure reflects that of the one in the main text.*

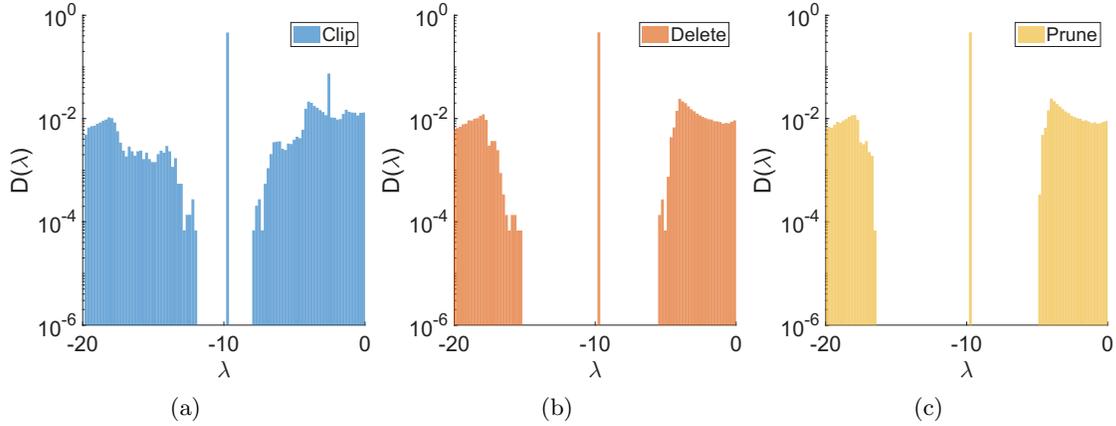

**Figure SM10.** *Density $D(\lambda)$ of eigenvalues $\lambda$ for equilateral metric networks generated from a Delaunay tessellation of a $\mathbb{Z}^2$ URL with $a = 0.1$ and (a) clip, (b) delete, and (c) prune network structure boundary conditions. The color of each figure reflects that of the one in the main text.*

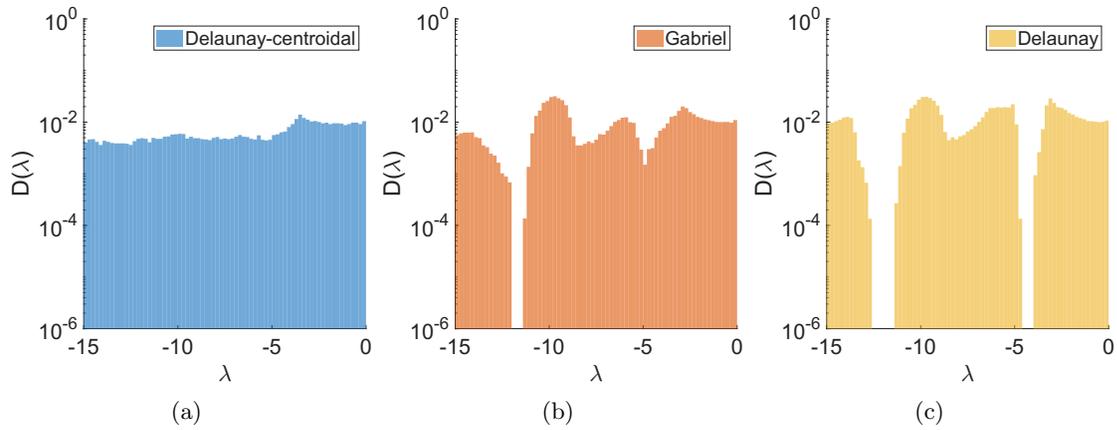

**Figure SM11.** Density $D(\lambda)$ of eigenvalues $\lambda$ for standard metric networks generated from a (a) Delaunay-Centroidal, (b) Gabriel, and (c) Delaunay tessellation of a $\mathbb{Z}^2$ URL with $a = 0.1$ and prune network structure boundary conditions. The color of each figure reflects that of the one in the main text.

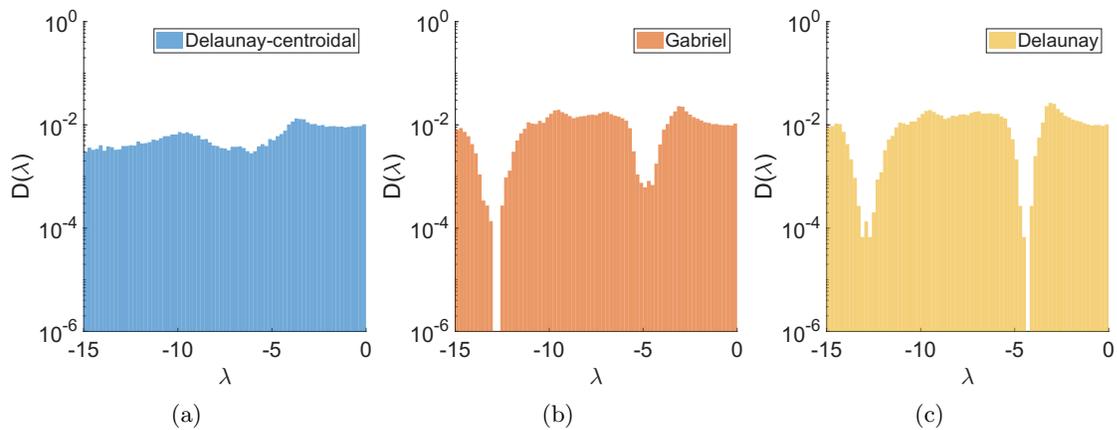

**Figure SM12.** Density $D(\lambda)$ of eigenvalues $\lambda$ for standard metric networks generated from a (a) Delaunay-Centroidal, (b) Gabriel, and (c) Delaunay tessellation of a disordered stealthy hyperuniform point pattern with $\chi = 0.48$ and prune network structure boundary conditions. The color of each figure reflects that of the one in the main text.

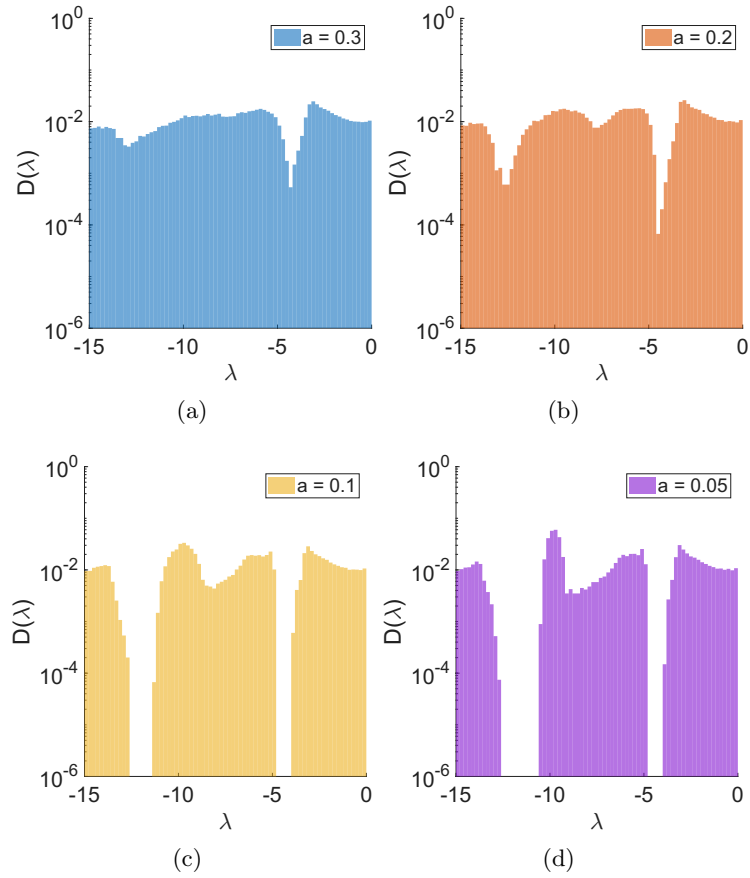

**Figure SM13.** *Density $D(\lambda)$ of eigenvalues $\lambda$ for standard metric networks generated from a Delaunay tessellation of a $\mathbb{Z}^2$ URL with $a =$ (a) 0.3, (b) 0.2, (c) 0.1, and (d) 0.05 and prune network structure boundary conditions. The color of each figure reflects that of the one in the main text.*

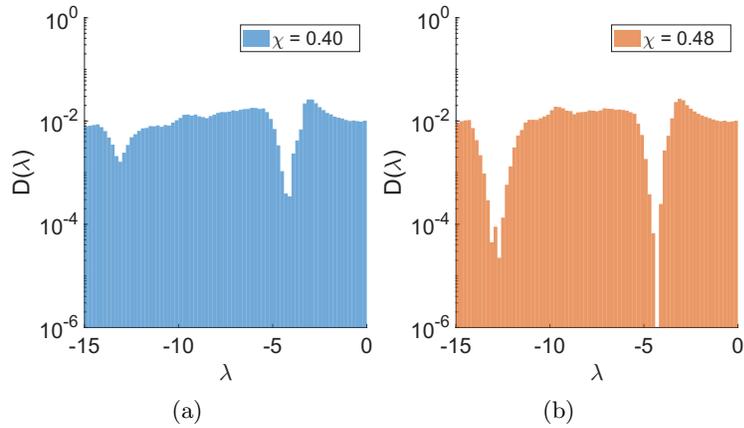

(a)    (b)

**Figure SM14.** Density $D(\lambda)$ of eigenvalues $\lambda$ for dtandard metric networks generated from a Delaunay tessellation of a disordered stealthy hyperuniform point pattern with $\chi =$ (a) 0.40 and (b) 0.48 and prune network structure boundary conditions. The color of each figure reflects that of the one in the main text.

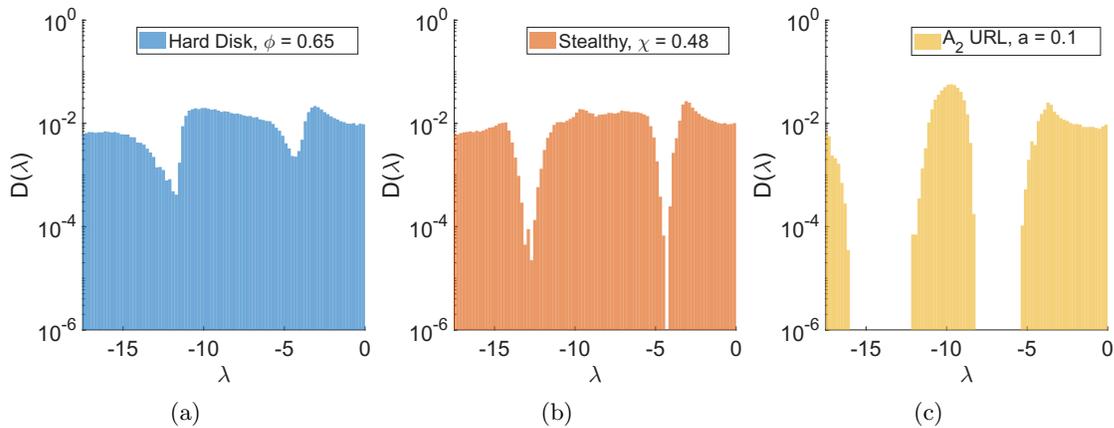

(a)    (b)    (c)

**Figure SM15.** Density $D(\lambda)$ of eigenvalues $\lambda$ for standard metric networks generated from a Delaunay tessellation of (a) an equilibrium hard disk fluid with $\phi = 0.65$, (b) a disordered stealthy hyperuniform point pattern with $\chi = 0.48$, and (c) an $A_2$ URL with $a = 0.1$ and prune network structure boundary conditions. The color of each figure reflects that of the one in the main text.

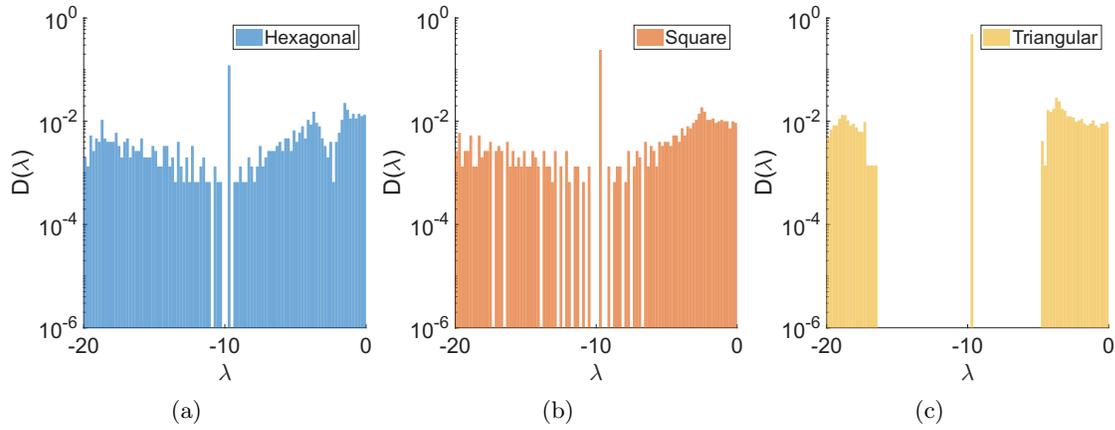

**Figure SM16.** *Density $D(\lambda)$ of eigenvalues $\lambda$ for equilateral metric networks generated from a finite portion of (a) a hexagonal tiling, (b) a square tiling, and (c) a triangular tiling. The color of each figure reflects that of the one in the main text.*

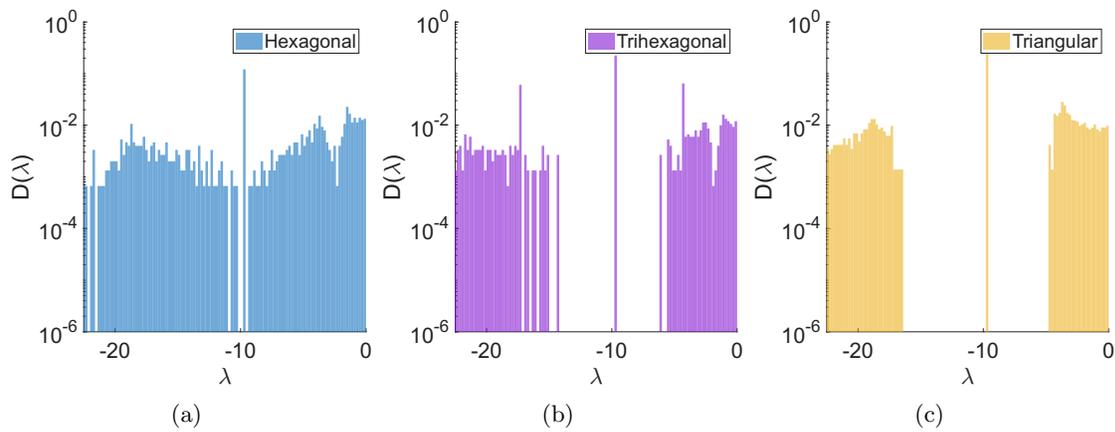

**Figure SM17.** *Density $D(\lambda)$ of eigenvalues $\lambda$ for equilateral metric networks generated from a finite portion of (a) a hexagonal tiling, (b) a trihexagonal (Kagome) tiling, and (c) a triangular tiling. The color of each figure reflects that of the one in the main text.*

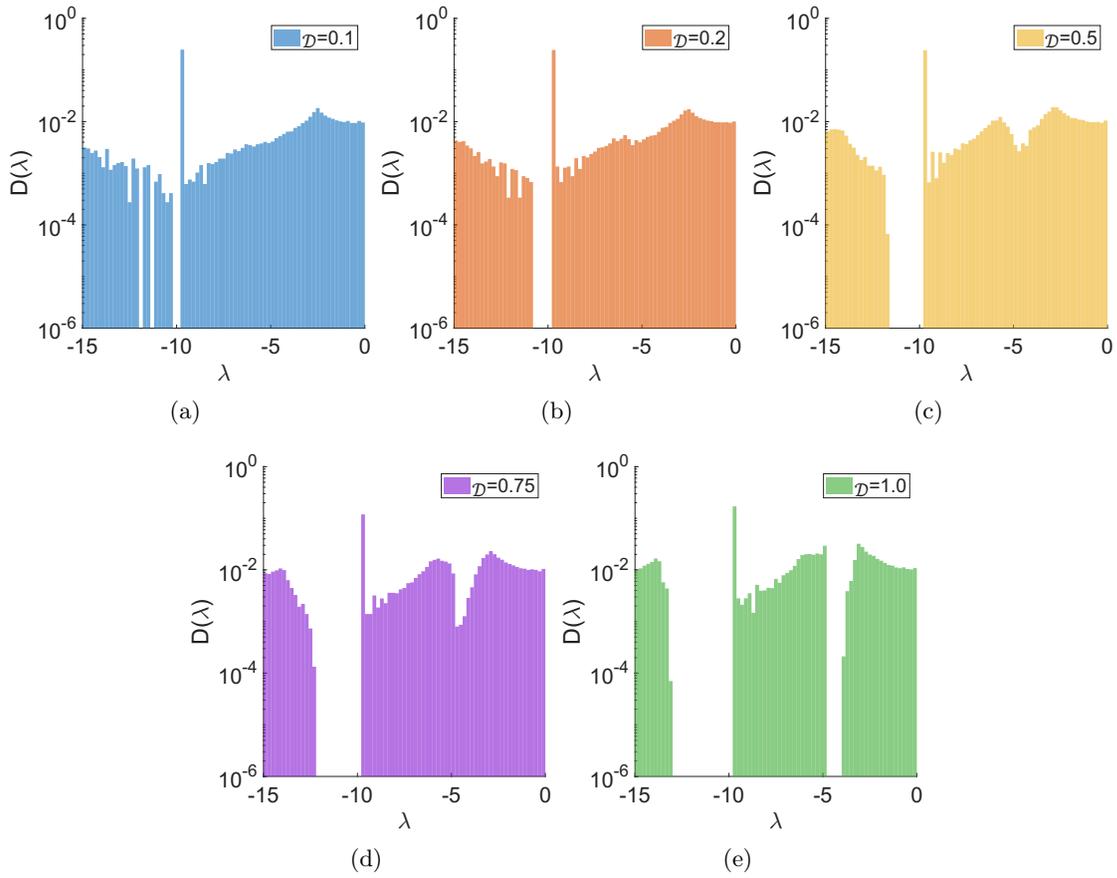

**Figure SM18.** *Density $D(\lambda)$ of eigenvalues $\lambda$ for standard metric networks generated from a finite portion of the square tiling with a fraction $\mathcal{D} =$ (a) 0.1, (b) 0.2, (c) 0.5, (d) 0.75, and (e) 1.0 of the squares containing a randomly oriented diagonal edge. The color of each figure reflects that of the one in the main text.*

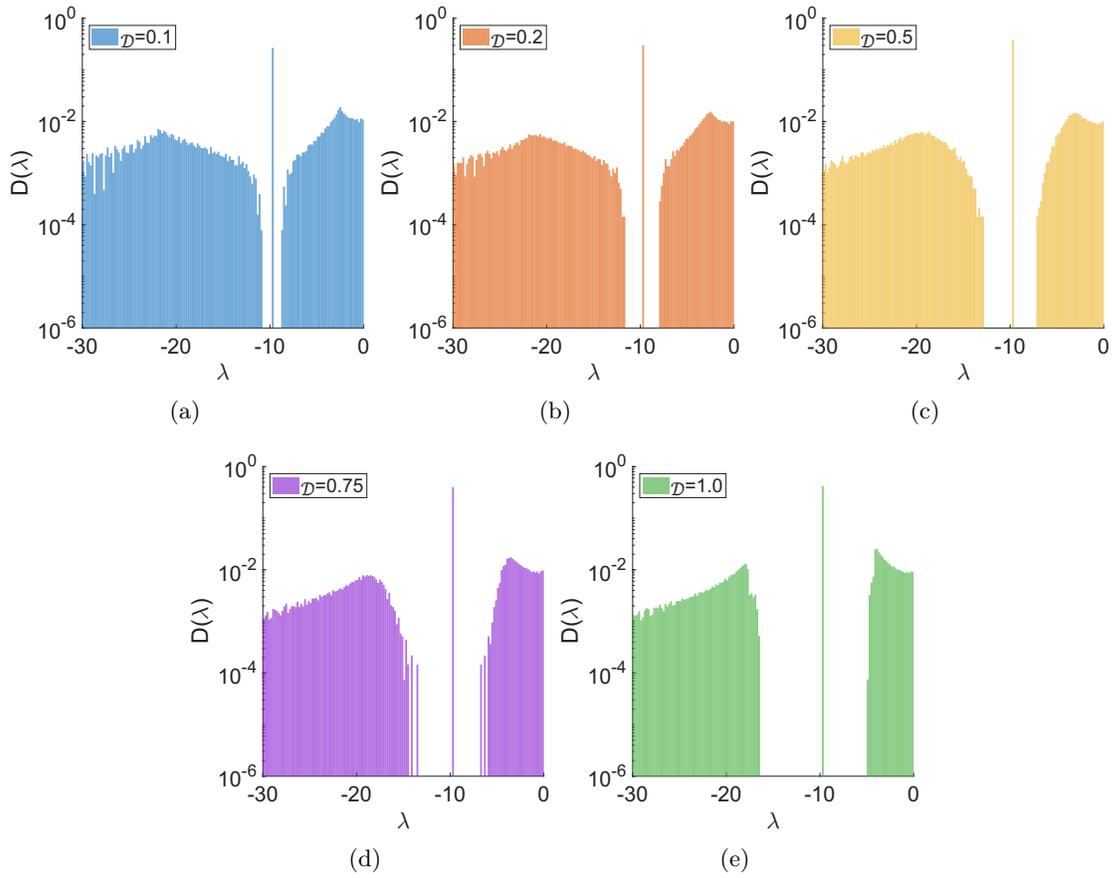

**Figure SM19.** *Density $D(\lambda)$ of eigenvalues $\lambda$ for equilateral metric networks generated from a finite portion of the square tiling with a fraction $\mathcal{D}$ = (a) 0.1, (b) 0.2, (c) 0.5, (d) 0.75, and (e) 1.0 of the squares containing a randomly oriented diagonal edge. The color of each figure reflects that of the one in the main text.*

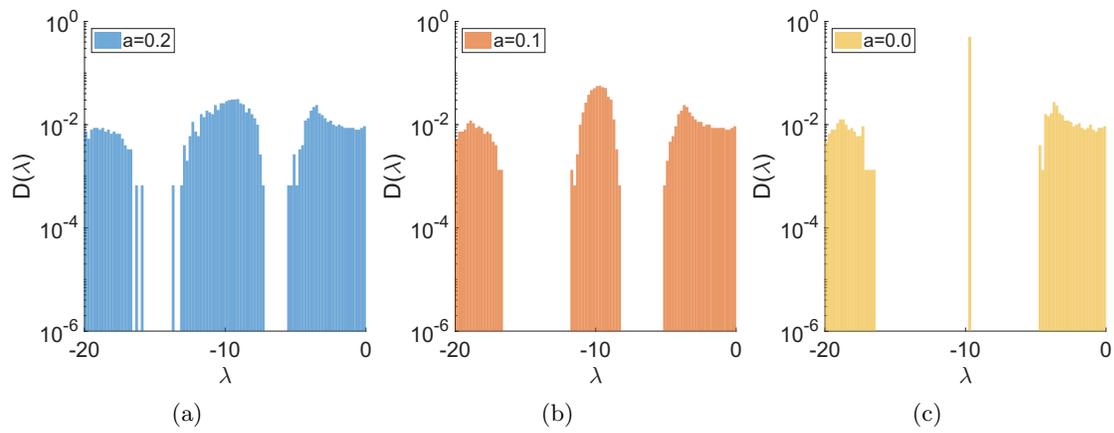

**Figure SM20.** *Density $D(\lambda)$ of eigenvalues $\lambda$ for standard metric networks generated from a Delaunay tessellation of a $\mathbb{A}_2$ URL with $a =$ (a) 0.2, (b) 0.1, and (c) 0.0 and prune network structure boundary conditions. The color of each figure reflects that of the one in the main text.*



\end{document}